\documentclass[apj]{emulateapj}
\usepackage{amssymb,amsmath}
\usepackage{mathrsfs}
\usepackage{natbib}
\usepackage{mathtools}
\usepackage{graphicx}
\usepackage{eucal}
\usepackage{url}
\usepackage{color}
\usepackage{float}
\usepackage{multirow}
\usepackage{pdfpages}
\usepackage[below]{placeins}

\usepackage{color}
\definecolor{red}{rgb}{1,0,0}

\DeclareMathOperator{\sech}{sech}

\begin{document}

\newcommand{\mt}{\ensuremath{\mathrm{M_\star }}}
\newcommand{\sfr}{\ensuremath{\mathrm{\dot{M}_\star }}}
\newcommand{\massunits}{\ensuremath{\mathrm{M_\odot }}}
\newcommand{\sfrunits}{\ensuremath{\mathrm{M_\odot}} \ensuremath{\mathrm{yr^{-1}}}}
\newcommand{\D}{\ensuremath{\mathcal{D}}}
\newcommand{\M}{\ensuremath{\mathcal{M}}}
\newcommand{\B}{\ensuremath{\mathcal{B}}}

\title{Sizing Up the Milky Way: A Bayesian Mixture Model Meta-Analysis of Photometric Scale Length Measurements}
\author{Timothy C. Licquia\altaffilmark{1,2}, Jeffrey A. Newman\altaffilmark{1,2}}
\altaffiltext{1}{Department of Physics and Astronomy, University of Pittsburgh, 3941 O'Hara Street, Pittsburgh, PA 15260; {tcl15@pitt.edu, janewman@pitt.edu} }
\altaffiltext{2}{Pittsburgh Particle physics, Astrophysics, and Cosmology Center (PITT PACC)}

\begin{abstract}
The exponential scale length ($L_d$) of the Milky Way's (MW's) disk is a critical parameter for describing the global physical size of our Galaxy, important both for interpreting other Galactic measurements and helping us to understand how our Galaxy fits into extragalactic contexts. Unfortunately, current estimates span a wide range of values and often are statistically incompatible with one another.  Here, we perform a Bayesian meta-analysis to determine an improved, aggregate estimate for $L_d$, utilizing a mixture-model approach to account for the possibility that any one measurement has not properly accounted for all statistical or systematic errors.  Within this machinery we explore a variety of ways of modeling the nature of problematic measurements, and then employ a Bayesian model averaging technique to derive net posterior distributions that incorporate any model-selection uncertainty.  Our meta-analysis combines 29 different (15 visible and 14 infrared) photometric measurements of $L_d$ available in the literature; these involve a broad assortment of observational datasets, MW models and assumptions, and methodologies, all tabulated herein.  Analyzing the visible and infrared measurements separately yields estimates for $L_d$ of $2.71^{+0.22}_{-0.20}$ kpc and $2.51^{+0.15}_{-0.13}$ kpc, respectively, whereas considering them all combined yields $2.64\pm0.13$ kpc.  The ratio between the visible and infrared scale lengths determined here is very similar to that measured in external spiral galaxies.  We use these results to update the model of the Galactic disk from our previous work, constraining its stellar mass to be $4.8^{+1.5}_{-1.1}\times10^{10}$ M$_\odot$, and the MW's total stellar mass to be $5.7^{+1.5}_{-1.1}\times10^{10}$ M$_\odot$.
\end{abstract}


\section{Introduction} \label{sec:intro}
Since the invention of the first telescopes, astronomers have been trying to explain the distribution of stars that make up the Milky Way (MW).  The earliest maps of our Galaxy were developed by keeping a simple tally of the number of stars one could see as a function of their apparent brightness and position on the sky and then interpreting these star counts with a few basic assumptions \citep{Herschel1785,Kap20,Kapteyn22,Seares25,Bok37,Oort38}.  Remarkably, while the accessibility and quality of data has been drastically transformed by advancing technology, this same basic methodology has underlain most present-day photometric models of the MW \citep{Bahcall86}, with only a small subset of recent studies that utilized more sophisticated techniques.  Today, a wealth of high-quality, well-calibrated observational data for stars has been accumulated from a wide array of multi-band photometric surveys that have been carried out over the past three decades, using both visible and infrared (IR) light.  As a result, the literature is rich with studies on the geometrical structure of the Galaxy.

The current picture of the MW has been radically transformed since the first pioneering papers, which followed the advent of detailed studies of extragalactic spiral galaxies \citep{deV59,Freeman70,Kormendy77}.  Today, it is well understood that the major stellar components of the MW include a bar with a bulge or pseudobulge at its center and a flattened disk that is much more visibly extended \citep[see, e.g.,][and references therein]{Licquia1}.  Generally, current models assume that the distribution of stars comprising the disk \emph{to first order} is axisymmetric and follows an exponentially declining density profile, both radially and vertically, such that the volume density may be written as
\begin{equation}
\rho_\star(R,\phi,Z) = \frac{\Sigma_\star(0)}{2H_d}\exp(-\frac{R}{L_d}-\frac{Z}{H_d}), \label{eq:rho}
\end{equation}
where $R$, $\phi$, and $Z$ are the Galactocentric cylindrical coordinates, $\Sigma_\star(0)$ is the central stellar surface density, $H_d$ is the (vertical) disk scale height, and $L_d$ is the (radial) disk scale length.  In some cases, authors alternatively employ an isothermal-sheet model for the vertical structure that replaces the $\exp(Z)$ dependence with a $\sech^2(Z)$ dependence in Equation \eqref{eq:rho} with the appropriate renomalization factors \citep[cf.][]{Spitzer42,vdKSearle,Freeman78}; however, the details of this are not of interest here.  More importantly, $L_d$ represents the radius containing the first $e$-folding of starlight within the disk in projection, or in other words where the surface density declines to $\sim$37\% of $\Sigma_\star(0)$, and hence provides a standard measure of the absolute physical size of the Galactic disk.

Dozens of attempts have been made to determine $L_d$ over the past few decades, making it one of the most investigated characteristics of our Galaxy.  Here, we focus exclusively on those estimates from photometric models of the MW in order to enable direct comparisons to measurements of extragalactic objects.  We have collected a total of 29 different measurements from the literature since 1990, 15 of them are based on optical data and 14 on IR data.  Altogether, these values lie in the range of $\sim$2--6 kpc, and upon close inspection reveal little consensus on the true size of the Galactic disk.  At least some of this disparity is likely due to variations in the assumptions that go into each MW model, which typically include between one and five stellar components that are fit by up to a dozen free parameters.  Other possible issues are that unidentified substructures present in the data are biasing models fit to particular lines of sight, or that there are a multitude of very different models that are roughly equally successful in fitting the data \citep{Juric08}.  To account for these complications many authors have incorporated substructure features into their models (e.g., spiral arms and rings), while others test a variety of functional forms for the assumed density law.

In this paper, we address the question: given the measurements available in the literature, what is the best photometric estimate of the MW's disk scale length?  In \citet[][in preparation]{Licquia3}, a companion paper to this study, we have found that scale lengths measured from optical photometry of other massive spiral galaxies \citep{Hall12}, which employed the same exponential density model as Equation \eqref{eq:rho}, span a range of $\sim$1--10 kpc.  This is rather comparable to the range of values for the MW described above.  To determine more precisely where the MW falls within this range, we can perform a Bayesian mixture model (BMM) meta-analysis \citep{Licquia1} of Galactic disk scale length estimates.  This method will enable us to investigate and remedy any sources of tension amongst disk scale length measurements.  Simultaneously, it will provide a single aggregate result that is built upon the rich assortment of photometric survey data that is available, but that also accounts for the possibility that any of the included estimates are offset due to systematics or bear an underestimated error bar, and then incorporate that information into the overall uncertainties in the combined result.  A BMM analysis will yield improved constraints on the Galactic $L_d$, which in turn will help us to better understand how our Galaxy measures up to its extragalactic peers.

The structure of this paper is as follows.  In \S\ref{sec:data}, we begin by describing the sample of Galactic $L_d$ estimates we have obtained, emphasizing the variety of observational data, MW models, and analysis techniques that they involve.  In \S\ref{sec:methods}, we explain the first-order corrections we make in order to place these estimates on an equal footing, and the BMM meta-analysis we subsequently perform on the resulting dataset.  Here, we also introduce a Bayesian model averaging technique that we use to produce our final posterior distributions and explain why it is appropriate to use in this work.  In \S\ref{sec:results} we present the aggregate results for $L_d$, including those from both segregating and combining the IR and visible data.  Here, we also present the results for parameters which characterize the overall consistency of the estimates in our dataset, as well as the different ways we have tested for robustness.  In \S\ref{sec:discussion}, we provide comparisons to $L_d$ estimates that have been determined from dynamical modeling, as well as to visible-to-IR scale length ratios measured for external galaxies.  In this section we also construct an updated model of the stellar disk using the results found here in order to revise our previous estimate of the total stellar mass from \citet{Licquia1}.  Lastly, in \S\ref{sec:summary} we summarize this study and highlight our conclusions.

\section{The Milky Way Disk Scale Length Dataset} \label{sec:data}
We have collected 29 different measurements of the exponential scale length, $L_d$, of the MW's disk published since 1990.  Generally speaking, each $L_d$ estimate is produced by fitting a model of the stellar components comprising the MW to observations of either resolved stars or unresolved starlight; 15 of these measurements were made in the visible and 14 in the infrared (IR).  We have restricted this study to include only measurements made based on visible and IR starlight to match as closely as possible what is done for external galaxies.  Hence, we have excluded dynamical estimates of the Galactic $L_d$ that are constrained by fitting MW \emph{mass} models to stellar kinematic data, as well those that critically rely on segregating stars into subpopulations based upon their spectroscopically-determined elemental abundances, as such measurements are infeasible for other galaxies.

In Figure \ref{fig:ld_evol} we display the values and time evolution of the $L_d$ estimates that we have collected, illustrating that as an ensemble they span $\sim$2--6 kpc, though many cluster in the 2--3 kpc range.  Taking these estimates at face value, we find that they display an interquartile range of 1.2 kpc and that their simple median value is $2.5\pm0.2$ kpc, which at first glance indicates fairly good overall agreement amongst them.  However, such metrics neglect the estimated errors in these measurements, not to mention the fact that several lack any error estimate at all (these are marked by dashed error bars in Figure \ref{fig:ld_evol}, which depict 25\% of the central value); hence, such a summary discards considerable amounts of information.  Furthermore, Figure \ref{fig:ld_evol} illustrates that many of the individual estimates are incompatible with one another at the $\geq$1$\sigma$ level, which indicates the possibility of systematic errors.

The presence of systematics would not be surprising; models of the MW are complex and require many assumptions about its various components.  Variations in these model choices amongst authors represents a major potential source of systematic uncertainty in the $L_d$ estimates we have collected.  As we will discuss in \S\ref{sec:methods}, utilizing a BMM meta-analysis technique, as opposed to more simple methods, to combine these estimates into a single aggregate result is useful on several fronts.  Most importantly, it allows us to produce posterior results that incorporate the possibility that any one measurement may be faulty, such as being systematically offset or having an overly-optimistic error estimate, due to not exploring the effects of varying model assumptions.  It also provides more flexibility, e.g., in dealing with those measurements with no formal error estimates, which likely deserve less credence, but still may contain valuable information about the true $L_d$ and hence should not be discarded entirely.  In cases where there are disagreements between measurements, the BMM technique used here will also degrade the uncertainties in derived parameters appropriately.
\begin{figure*}[T]
%\begin{figure}[T]
\centering
\includegraphics[page=1,trim=.7in .75in .8in .65in, clip=true, width=\textwidth]{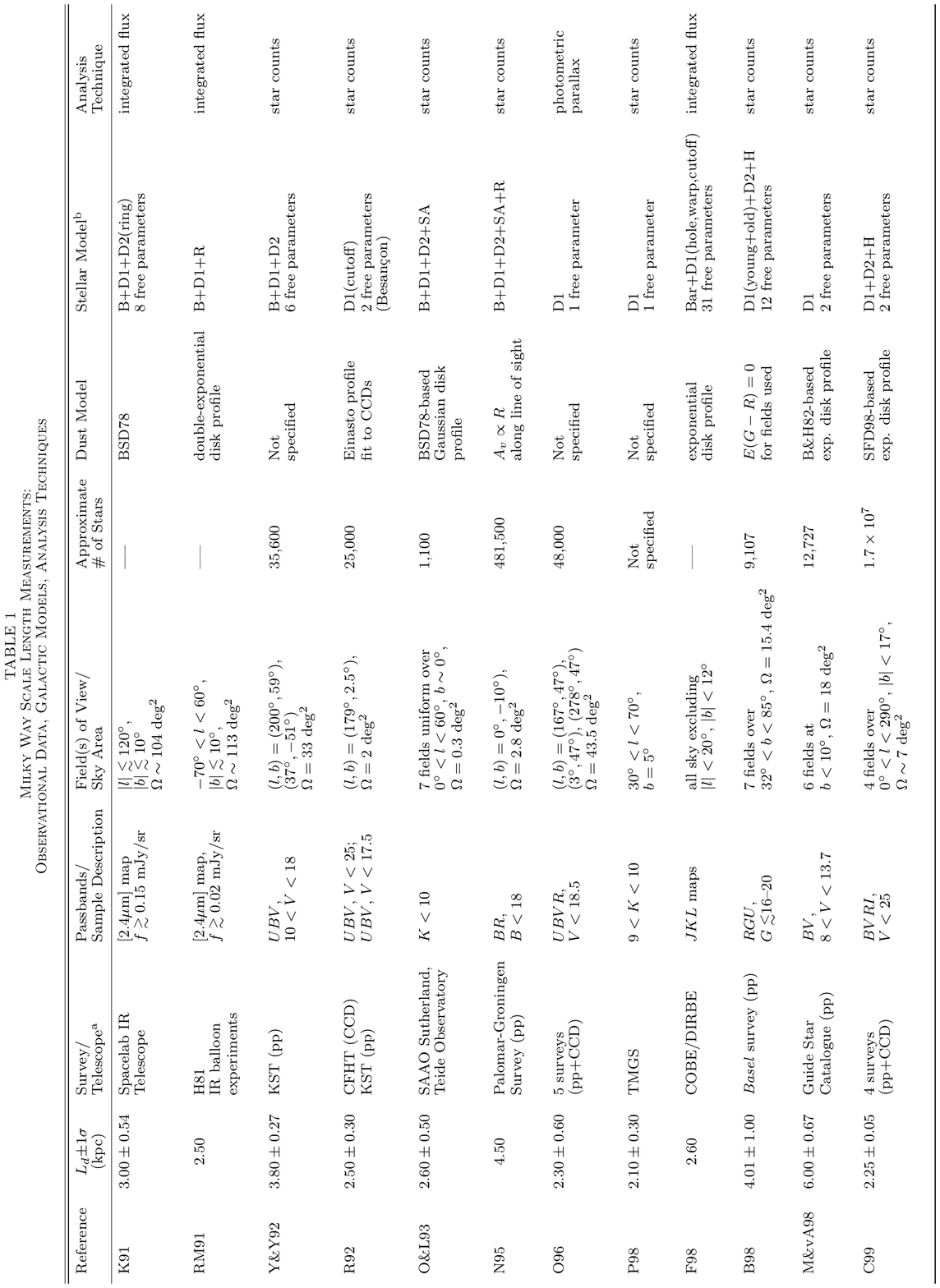}
%\end{figure}
\end{figure*}
\begin{figure*}[h!]
%\begin{figure}[h!]
\centering
\includegraphics[page=2,trim=.8in .75in .7in .65in, clip=true, width=\textwidth]{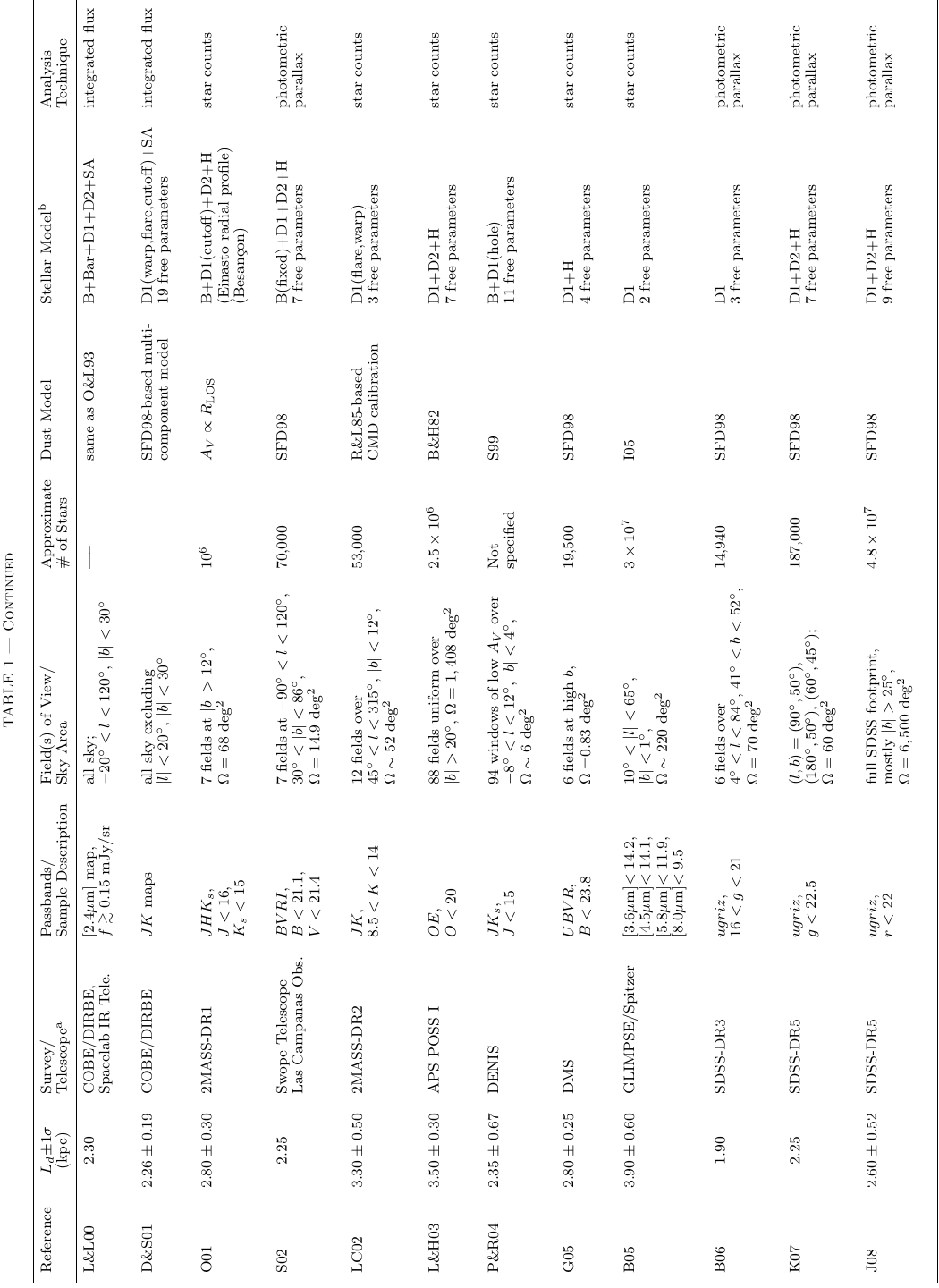}
\vspace{1in}
%\end{figure}
\end{figure*}
\begin{figure*}[h]
%\begin{figure}[h!]
\centering
\includegraphics[page=3,trim=0.8in .75in .7in .65in, clip=true, width=\textwidth]{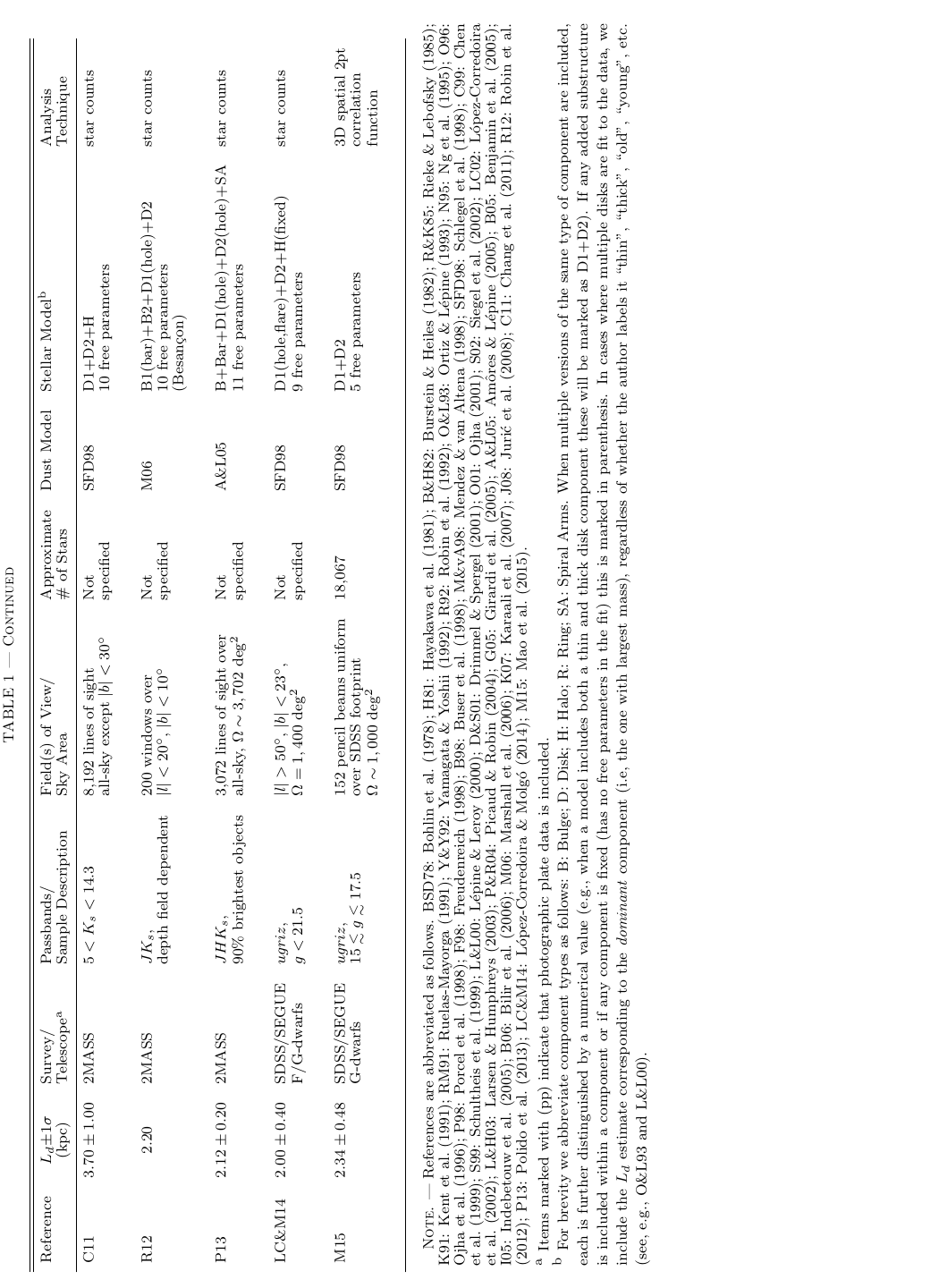}
%\end{figure}
\end{figure*}
\clearpage

The studies that we have collected employ a rich assortment of MW structural models, observational datasets, and analysis techniques.  In the following subsections, we detail the diversity in each of these three ingredients that go into producing each $L_d$ estimate that we have included in our analyses.  We have also summarized these details in Table 1 for ease of comparison.  As noted in the Table 1 footnote, hereafter we will use the following reference abbreviations.  BSD78: \citet{BSD78}; H81: \citet{Hayakawa81}; B\&H82: \citet{BH82}; R\&K85: \citet{RK85}; K91: \citet{Kent91}; RM91: \citet{Ruelas-Mayorga91}; Y\&Y92: \citet{YamagataYoshii92}; R92: \citet{Robin92}; O\&L93: \citet{OrtizLepine93}; N95: \citet{Ng95}; O96: \citet{Ojha96}; P98: \citet{Porcel98}; F98: \citet{Freud98}; B98: \citet{Buser98}; M\&vA98: \citet{MendezvanAltena98}; SFD98: \citet{SFD98}; C99: \citet{Chen99}; S99: \citet{Schultheis99}; L\&L00: \citet{LepineLeroy00}; D\&S01: \citet{DrimmelSpergel01}; O01: \citet{Ojha01}; S02: \citet{Siegel02}; LC02: \citet{Lopez02}; L\&H03: \citet{LarsenHumphreys03}; P\&R04: \citet{PicaudRobin04}; G05: \citet{Girardi05}; A\&L05: \citet{Amores05}; B05: \citet{Benjamin05}; I05: \citet{Indebetouw05}; B06: \citet{Bilir06}; M06: \citet{Marshall06}; K07: \citet{Karaali07}; J08: \citet{Juric08}; C11: \citet{Chang11}; R12: \citet{Robin12}; P13: \citet{Polido13}; LC\&M14: \citet{Lopez14}; M15: \citet{Mao15}.

\subsection{Observational Data} \label{sec:data_obs}
In this study we are interested only in photometric measurements of the MW's scale length, which we wish to match \emph{as closely as possible} the methods used in measurements for other disk galaxies.  Therefore, the observational data employed by our set of studies is strictly limited to measurements of starlight at either visible or IR wavelengths.  Despite this limitation, there are many Galactic surveys that fit this requirement; notable examples are the Two Micron All-Sky Survey (2MASS) in the IR and the Sloan Digital Sky Survey (SDSS) in the visible.  While these large-scale surveys covered a large fraction of the night sky, many more smaller-scale projects were carried out to yield pencil-beam surveys of stars in the Galaxy, encompassing many different fields of view with varying sizes.
%\begin{figure*}[ht]
\begin{figure}[t]
\centering
\includegraphics[width=\columnwidth, trim=.35in .1in .65in .5in, clip=true]{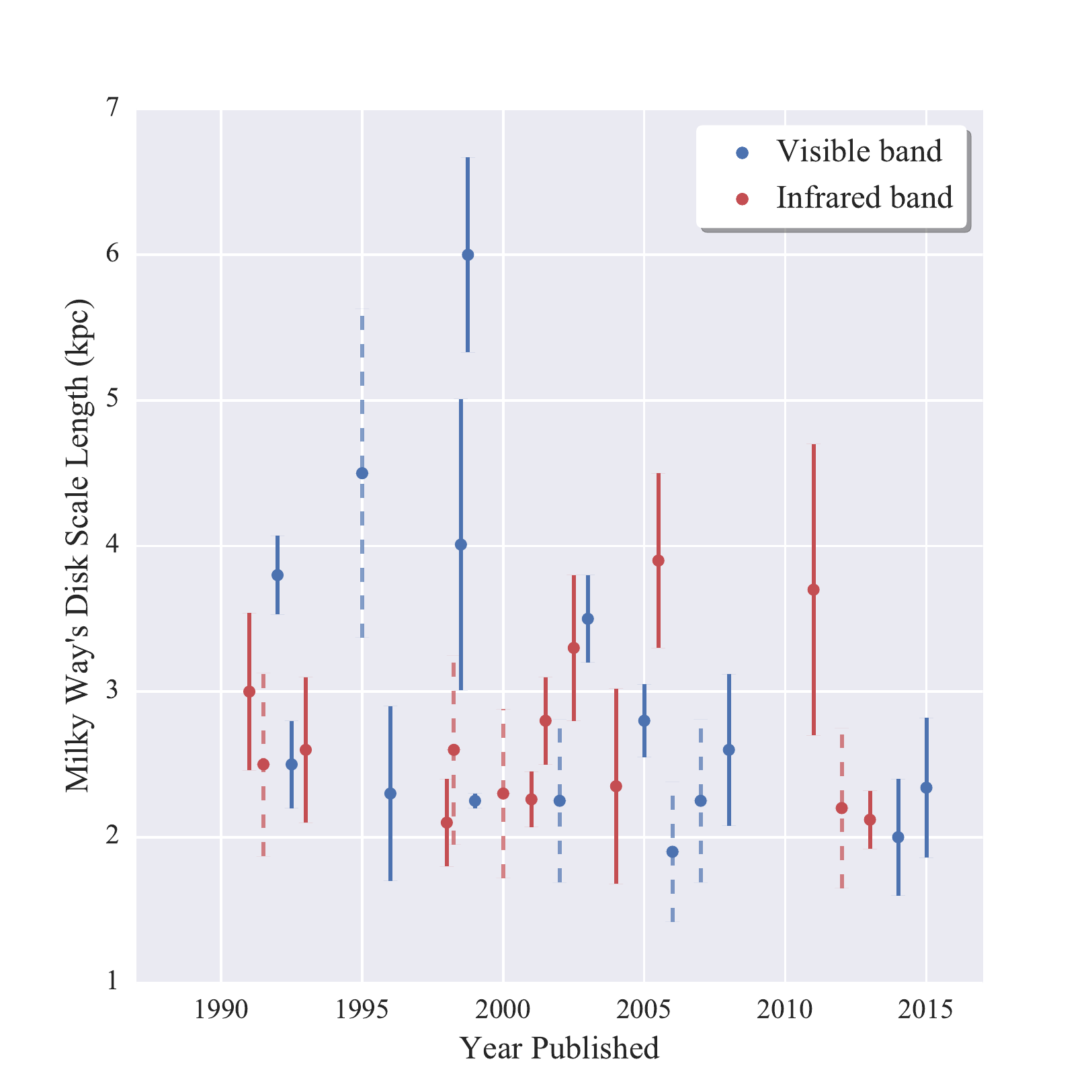}
\caption{The evolution of Milky Way's $L_d$ estimates in the literature since 1990, which span $\sim$2--6 kpc.  There are 29 measurements in total; 15 are determined from visible starlight (shown in blue) and 14 from infrared starlight (shown in red).  For those measurements that lack any error estimate, we display dashed error bars that correspond to 25\% of the central value, which will be the standard treatment for our analyses (see \S\ref{sec:corrections}).  A detailed list of the studies that produced these estimates is provided in Table 1.}
\label{fig:ld_evol}
\end{figure}
%\end{figure*}

Earlier studies using measurements made in the visible were generally limited to high Galactic latitude, $b$, in order to minimize reddening and extinction due to dust (e.g., Y\&Y92; O96; B98), or to narrow windows of low absorption (e.g., R92).  This limitation was mitigated, however, by the all-sky map of Galactic dust provided by SFD98, allowing for measurements based on optical data to push toward lower $b$ (e.g., C99; S02; J08).  In some cases, authors use the full extinction correction out to infinity as a boundary condition for their dust model (see, e.g., G05, which employs an exponential dust distribution).  We note here that many of the more recent studies that utilize the SFD98 dust maps, particularly those based on SDSS data, generally apply the full extinction correction out to infinity for each star in their sample.  This clearly is an overcorrection and will bias distance or magnitude estimates; however, J08 explore the effects of this and find that it is only worrisome for stars within 100 pc of the Galactic plane when fitting structure models.  For their case, this corresponds to only 0.05\% of their stellar sample (which they exclude) given the magnitude limits of the SDSS instrument, concluding that overestimated interstellar extinction corrections will have negligible impact on results.  IR measurements, on the other hand, are far more immune to the effects of interstellar dust, allowing studies to probe the Galactic structure over a wide range in $b$ (e.g., K91; D\&S01; B05).

The pertinent details of each observational dataset are listed in Table 1.  There one can see that fields of view can range from $\sim$1 deg$^2$ to covering the full sky, corresponding to stellar sample sizes that vary from $\sim$10$^3$ to $\sim$10$^7$.  We note that the cutoff date of 1990 is used to limit ourselves primarily to observational data yielded from CCD technology, though a few of our $L_d$ estimates are derived from observations using photographic plates (as is noted in Table 1), as well as to ensure the use of modern Galactic assumptions.  We note that in some cases the IR data employed did not contain a sample of photometrically resolved stars, but instead provided a map of integrated flux from Galactic light.

\subsection{Models of the Milky Way} \label{sec:data_models}
Almost all of the studies we have collected model the radial structure of the Galactic disk according to Equation \eqref{eq:rho}.  The exceptions to this are R92 and O01 who use an \citet{Einasto65} law, which yields a radial density distribution that is equivalent to an exponential profile and hence yields comparable estimates of $L_d$.

Depending on the field of view and depth of the survey, other components of the Galaxy may be included as well.  For example, a particular model may include a bulge, a second disk, or a halo, each with their own density profile.  However, with this added level of model complexity comes additional free parameters that need to be fitted for simultaneously.  As a result, strong inter-parameter degeneracies can often hamper the ability to isolate a single ``correct'' description of the data, which is likely the main culprit for the disparate values of $L_d$ that can be found in Table 1.

Lastly, there are various added features and/or substructure that are often included in the disk models employed.  J08 pointed out that the presence of unidentified substructures in the true distribution of stars can bias the results from models attempting to fit the smooth, underlying profile.  To handle this, many studies include substructure components in their Galactic model, such as rings and/or spiral arms (e.g., O\&L93; N95; D\&S01; P13).  Others also allow for the possibility of an asymmetry or ``warp'' in the shape of the disk (e.g., LC\&M14).  Alternatively, some of the included studies may allow or even force their model to deviate from a pure exponential profile.  This can take the form of truncations or holes at the center of the Galactic disk, which systematically ascribe more stellar material to the bulge component in this region (e.g., F98; P\&R04; LC\&M14).  Others include a cutoff radius beyond which they assume stars cease to exist, instead of extrapolating to $R=\infty$ (e.g., F98).  All of these assumptions are plausible; other spiral galaxies are observed both with and without central holes and rings, their profiles can appear either flocculent or smooth, as well as warped or flat \citep{Buta14,Buta15}.

\subsection{Analysis Techniques Used for Scale Length Measurements} \label{sec:data_fitting}
As seen in Table 1, even when similar sorts of observational data are used, a wide variety of analysis techniques have been employed to measure the MW's disk scale length.  We provide a summary of these methods here.

\subsubsection{Integrated Light}
The first case to consider is where the observational data take the form of a smooth map of the integrated flux from Galactic light, typically produced by IR instruments with lower angular resolution.  This provides the intensity of emission at some effective frequency $\nu$, $I_\nu$, as a function of Galactic longitude and latitude, $(l,b)$.  To model $I_\nu$, one first needs to adopt a model for the distribution of stars in the MW, $\rho_\star(R,\phi,Z)$, where the disk component typically follows the form of Equation \eqref{eq:rho}.  By assuming values for the Galactocentric radius and planar offset of the Sun, $R_0$ and $Z_0$, respectively, one can then recast $\rho_\star$ into heliocentric spherical coordinates $(r,l,b)$ by
\begin{equation}
R=(R_0^2 + r^2 \cos^2b - 2 R_0r \cos b \cos l)^{1/2}, \nonumber
\end{equation}
\begin{equation}\label{eq:conversion}
Z = r \sin b + Z_0,
\end{equation}
where $r$ is the line-of-sight distance toward $(l,b)$.  Next, one assumes a characteristic specific emissivity, $\epsilon_\nu$, for the stellar population (equivalent to assuming a luminosity function), which is multiplied by $\rho_\star$ to obtain the stellar flux density.  The observed integrated flux will be modulated by dust absorption, however, and so one must adopt a model for the distribution of dust in order to obtain the optical depth along the line of sight, $\tau_\nu(r)$.  Finally, all of these ingredients are combined into a model of the light intensity across the sky by
\begin{equation}
I_\nu(l,b) = \sum_{c} \epsilon_{\nu,c} \int_{0}^{\infty} \rho_{\star,c}(r,l,b) e^{-\tau_\nu(r)} \,\text{d}r,
\end{equation}
where the summation over $c$ signifies the different stellar components and populations that need to be included (see \S3 of D\&S01 for more details).  One then optimizes the free parameters included in the model to best fit the observational data, which yields an estimate of $L_d$.

\subsubsection{Star Counts}
While star count analyses using photometric data began nearly a century ago \citep[as described in][]{Bahcall86,BM98}, the practicality and hence the popularity of this technique for studying the geometric structure and size of the MW has exploded over the past few decades (see the final column of Table 1).  This is not only due to the accuracy and depth of photometric surveys improving with technology, but also because imaging of extragalactic objects has resulted in better-informed models of stellar structure \citep[e.g.,][]{dVP78,BS1,BS2}.  Star count analyses have been the most prevalent method for constraining photometric models of the MW.

Here, one attempts to reproduce the observed number of stars in a field of view subtending solid angle $\Omega$ toward the direction $(l,b)$ whose apparent magnitudes are between some lower and upper limit, denoted by $m_1$ and $m_2$, respectively.  In order to do so, one must integrate the assumed stellar density law, $\rho_\star(R,\phi,Z)$, weighted by an assumed luminosity function, $\Phi(M)$, over the volume of the survey and the range of magnitudes.  After converting $\rho_\star$ to heliocentric spherical coordinates via Equation \eqref{eq:conversion}, one obtains the ``fundamental equation of stellar statistics'' \citep{Bahcall86}:
\begin{multline}
N(m_1,m_2,l,b,\Omega) = \\
\quad \sum_{c}\Omega\int_{m_1}^{m_2} \int_{0}^{\infty} \rho_{\star,c}(r,l,b) \Phi_c(M) r^2  \,\text{d}r \,\text{d}m, \label{eq:fund_stellar_stats}
\end{multline}
where again the summation over $c$ implies that the individual stellar components are combined to yield the total.  Here, one also substitutes the absolute magnitude as $M=m-5\log r-A(r)+5$, where $A(r)$ is the total extinction out to distance $r$ in the passband that the magnitudes are measured using, making the above integrand a function purely of $r$ and $m$.  Therefore, one must also include a model of Galactic dust and its attenuation to inform the choice of $A(r)$.  Using Equation \eqref{eq:fund_stellar_stats}, authors optimize the free parameters included in $\rho_\star$ to achieve the best fit to the observed star counts, yielding an estimate of $L_d$.

The star counts method is often considered superior for making determinations of the 3D structural parameters of our Galaxy compared to the integrated-light technique described above.  As P\&R04 pointed out, integrated flux profiles are dominated by the brightest and closest stars, whereas star count studies incorporate a wider range of intrinsic luminosities and distances.  Furthermore, a common simplification strategy is to restrict to a narrow bin in color, allowing one to assume that all stars in the sample have approximately the same color and hence absolute magnitude via a Hertzsprung-Russell-diagram (HRD) relation, and essentially rendering $\Phi$ into a Dirac delta function (see, e.g., LC\&M14).  As P13 pointed out, studies using the star counts method can also be separated into two classes: those that adopt an empirical luminosity function (e.g., O\&L93; B98) and those that adopt a theoretical luminosity function based on models of stellar evolution (e.g., P\&R04; G05).

\subsubsection{Photometric Parallax}
Taking the star counts method one step further, authors have been able to make explicit distance measurements for each star in their dataset using a so-called ``photometric parallax'' relation (see, e.g., \citealp{GilmoreReid83}, O96, J08).  Here, one uses an appropriately-calibrated HRD to map each star's color to its intrinsic luminosity and combine that with its apparent magnitude in order to infer its distance, $r$.  Such a technique yields the full 3D position information $(r,l,b)$ for each star in a sample.  Authors then simply need to optimize the free parameters in their Galactic model $\rho_\star$ to best reproduce to volume density structure displayed by the sample, yielding an estimate for $L_d$.  

\subsubsection{Two-Point Correlation Function}
Recently, M15 have pioneered the application of the two-point correlation function, $\xi(r)$, in fitting the stellar density distribution.  Traditionally, this technique has been used to study the large-scale structure of the Universe, where $\xi(r)$ quantifies the strength of galaxy clustering as a function of separation \citep{Peebles80}.   This is calculated as
\begin{equation}
\xi(r) = \frac{DD(r)}{RR(r)}-1, \label{ref:tpcf}
\end{equation}
where $DD(r)$ is the observed number of pairs of galaxies at separations between $r$ and $r+\text{d}r$, and $RR(r)$ is the number of pairs measured in a mock catalog of galaxies randomly distributed over the same volume.  In order to constrain Galactic structure, M15 have replaced the idea of a \emph{random} catalog with a \emph{model} catalog, where the positions of stars are generated from their Galactic stellar density model, which will have counts of pairs $MM(r)$ as a function of separation.  Here, $DD(r)$ becomes the number of pairs of stars as a function of separation, and $MM(r)$ replaces $RR(r)$ in Equation \eqref{ref:tpcf}.  The free parameters of the model are then optimized to yield $\xi(r)\sim0$, producing an estimate of $L_d$.

We point the reader to Table 1, where we have provided a detailed synopsis for each of the published works that produced an $L_d$ estimate that we utilize in this study.  Here one can find a breakdown of the observational data and methodologies that go into each.  Where provided, this includes: the reference, the scale length estimate value, the survey/telescope utilized, the passbands that measurements were made using and any associated sample-defining criteria, the field(s) of view and/or sky area covered, the approximate number of stars in the sample, details of the authors' choice of dust and stellar models for the Galaxy, including both the major and substructure components incorporated and the corresponding number of free parameters, and lastly the analysis technique.

\section{Methods} \label{sec:methods}

\subsection{Bayesian Mixture-Model Technique} \label{sec:hb}
Our goal is to perform a meta-analysis of the $L_d$ estimates we have collected in order to determine a single aggregate result, encapsulating our overall knowledge from the various observational data and MW models available.  As discussed above, these estimates often are in tension with each other; this is likely due at least in part to differences in assumptions and methodologies amongst them.  We therefore rely on the power of a Bayesian mixture-model technique \citep{Press97, Hogg10, Dahlen13, Licquia1} to perform a meta-analysis of the set of literature results.  This is the appropriate tool to use for a variety of reasons.  Most importantly, it has been vetted as a powerful technique for extracting the consistent underlying signal from a sample of potentially flawed measurements \citep[see, e.g., ][]{LangHogg}.  It also yields error estimates that expand appropriately when the individual estimates are statistically inconsistent.  This is a distinct advantage over the inverse variance-weighted mean (IVWM), which has more limitations than the Bayesian framework; e.g., calculating the IVWM requires that the individual measurements are well-described by Gaussian probability distributions and that they are statistically compatible with one another.  However, one helpful feature of the Bayesian mixture-model technique is that it becomes equivalent to calculating the IVWM in the limit that the data that are all found to be highly consistent with one another.

In this study, we will produce our results using the same formalism and by employing the same problematic-measurement models that are laid out in \citet{Licquia1}, though with an added level of analysis that will be introduced in \S\ref{sec:BMA}; we refer the reader to this work for the full details, where one can also find a pedagogical introduction.  We note that in our previous work we have labeled this a ``hierarchical Bayesian technique,'' in the sense that our method relies on fitting for free parameters that characterize a subset of the data that we analyze simultaneous to the physical parameter of interest (e.g., $L_d$).  This is not to be confused with methods that recently have been labeled as ``Bayesian hierarchical modeling'' or ``Hierarchical Bayes'' \citep[e.g.,][]{Loredo12,Gelman13,Martinez15}, which are defined by having several layers of parameters, typically such that priors on model parameters are themselves dependent on additional free parameters (i.e., hyperparameters), and hence require their own priors (i.e., hyperpriors).  Therefore, to avoid confusion in the usage of ``hierarchical'', we will avoid using that terminology here, and instead refer to our method as a Bayesian mixture model (BMM) meta-analysis hereafter.  We will now briefly summarize the details of this method, which will be important for interpreting our tables and figures below.

In short, we begin by assuming that our $L_d$ dataset can be divided into two subgroups: one where measurements are constraining the same parameter of interest with properly-estimated uncertainties (``good''), and one where measurements are inaccurate in some way, e.g., suffering from unrealistic error estimates or neglecting a potentially significant source of bias (``bad'').  The probability distribution for the true value of $L_d$ given any single measurement then looks like $P(L_d) = f_\text{good} P({L_d \mid \text{measurement is ``good''}}) + (1 - f_\text{good}) P({L_d \mid \text{measurement is ``bad''}})$, where $P({A\mid B})$ generically indicates the probability distribution function (PDF) for $A$ conditional upon $B$ being true and $f_\text{good}$ denotes the probability that the measurement is ``good''.  This provides a template for the likelihood function within our BMM analysis.  If the measurement is represented by a Gaussian distribution with mean $\mu$ and standard deviation $\sigma$, then we can straightforwardly write that 
\begin{multline}
P({L_d \mid \text{measurement is ``good''}}) =  \\ 
\frac{1}{\sqrt{2 \pi \sigma^2}}\exp\left(\frac{-(L_d-\mu)^2}{2\sigma^2}\right).  \label{eq:Pgood} 
\end{multline}  
%\begin{equation} \label{eq:Pgood} P({L_d \mid \text{measurement is ``good''}}) = \frac{1}{\sqrt{2 \pi \sigma^2}}\exp\left(\frac{-(L_d-\mu)^2}{2\sigma^2}\right). \end{equation}  

As mentioned above, there are a variety of ways that measurements can be problematic.  We have developed a number of bad-measurement models to account for this; specifically, these are:
\begin{description}
  \item[free-$n$] we assume that bad measurements have overly-optimistic error bars by a factor of $n$.  Here $P({L_d \mid \text{measurement is ``bad''}})$ would be constructed by replacing $\sigma$ with $n\sigma$ in Equation \eqref{eq:Pgood};
  \item[free-$Q$] we assume that bad measurements have neglected a significant source of uncertainty, whose overall magnitude should be added in quadrature to the nominal error bars.  We quantify this as some fraction $Q$ of the median estimate from the set of measurements, which we will denote as $\mu^\text{MED}$, and construct $P({L_d \mid \text{measurement is ``bad''}})$ by replacing $\sigma^2$ with $\sigma^2 + (Q\mu^\text{MED})^2$ in Equation \eqref{eq:Pgood};
  \item[free-$F$] we assume that all measurements are truly capable of reaching only a certain level of accuracy, and bad measurements are those that fall below this level.  We quantify this by setting a floor value on error bars for all measurements uniformly equal to some fraction $F$ of $\mu^\text{MED}$, and hence we construct $P({L_d \mid \text{measurement is ``bad''}})$ by replacing $\sigma$ with $F\mu^\text{MED}$ in Equation \eqref{eq:Pgood} in all cases where $\sigma$ is below that limit;
  \item[$P_\text{bad}$-flat] we assume that bad measurements are critically flawed and should contribute zero weight to our study.  We achieve this by modeling $P({L_d \mid \text{measurement is ``bad''}})$ as a flat PDF over all parameter space.
\end{description}
Equations (2)--(5) in \citet{Licquia1} show the full form of the likelihood function for each case.  We note here that making first-order corrections to our dataset, which will be discussed in following subsections, can cause $\mu^\text{MED}$ to shift in value.  To circumvent any difficulties in interpreting our results due to this, throughout this work we invariably set $\mu^\text{MED} = 2.5$ kpc, as this very nearly if not equal to the median estimate value among the cases of considering the visible or IR measurements separately and when combining them.

Hereafter, we generically denote each model by $\M_k$ and its set of free parameters (separate from $L_d$) by $\Theta_k$; in most cases, those free parameters are $f_\text{good}$ and one of $n$, $Q$, or $F$.  Amongst all of these scenarios, $\Theta_k$ represents free parameters that characterize the data itself, which are fit for simultaneously with the true value of $L_d$.  We also explore the effect of setting $f_\text{good}=0$ within each of the models described above (except for the $P_\text{bad}$-flat model, as that would mean nullifying all of our data), which corresponds to assuming that not just \emph{some}, but all $L_d$ measurements are flawed to some extent; these cases are denoted by appending ``all-bad'' to the model name.  Lastly, we explore the possibility that all $L_d$ estimates are accurate by setting $f_\text{good}=1$ within any model that we have described above, which we denote as the ``all-good'' model.  As mentioned above, this reduces our formalism to calculating the IVWM, and serves as our null hypothesis.  Overall, this constitutes eight different ways of modeling the $L_d$ dataset; we will refer to this dataset as an ensemble with the label \D.

Assuming that all of the data are statistically independent, one can write the overall likelihood function as the product of the likelihood functions for each measurement,
\begin{equation} \label{eq:likelihood}
P({\D \mid L_d, \Theta_k, \M_k})  = \displaystyle\prod\limits_{i} P({\mu_i, \sigma_i \mid L_d, \Theta_k, \M_k}),
\end{equation}
where $\mu_i$ and $\sigma_i$ are the mean and error estimate for the $i$th estimate.  At this point, $f_\text{good}$ (which is buried in $\Theta_k$) may be interpreted as the fraction (or frequency) of accurate estimates in \D.  Finally, the joint posterior for the parameters of the mixture model given the set of measurements used can be calculated by
\begin{multline}
P({L_d, \Theta_k \mid \D, \M_k}) = \\
\qquad\frac{P({\D \mid L_d, \Theta_k, \M_k}) P(L_d) P({\Theta_k \mid \M_k})}{P({\D \mid \M_k})}.  \label{eq:posterior}
\end{multline}
%\begin{equation} \label{eq:posterior}
%P({L_d, \Theta_k \mid \D, \M_k}) = \frac{P({\D \mid L_d, \Theta_k, \M_k}) P(L_d) P({\Theta_k \mid \M_k})}{P({\D \mid \M_k})}.
%\end{equation}

\subsubsection{Posterior Results From Bayesian Model Averaging} \label{sec:BMA}
In \citet{Licquia1}, we compared the Bayesian evidence for each model to select the one that best fits the data.  For model $\M_k$ with free parameters contained in the vector $\Theta_k$, the evidence represents the probability of measuring the data \D\ given that $\M_k$ is the correct model and is calculated by
\begin{multline}
P({\D \mid \M_k}) = \\ \int\int P({\D \mid L_d, \Theta_k, \M_k})P(L_d)P({\Theta_k \mid \M_k}) \,\text{d}L_d\,\text{d}\Theta_k. \label{eq:evidence}
\end{multline}
%\begin{equation}
%P({\D \mid \M_k}) = \int\int P({\D \mid L_d, \Theta_k, \M_k})P(L_d)P({\Theta_k \mid \M_k}) \,\text{d}L_d\,\text{d}\Theta_k. \label{eq:evidence}
%\end{equation}
This is the integral of the likelihood weighted by the priors over all parameter space.  Such a metric provides a natural, Bayesian method of applying the principles of Occam's Razor to model selection by weighing the goodness-of-fit against the size of the parameter space required to achieve it.  In that study, we found that the all-good model came out well ahead in this comparison, such that the other models were not comparable to it.

Unlike the results in \citet{Licquia1}, we find here that in some cases more than one bad-measurement model can be competitive in maximizing the Bayesian evidence \emph{while also} yielding measurably different $P({L_d \mid \D, \M_k})$.  While these differences are well below the $\sim$1$\sigma$ threshold, there still remains the question of how to select the correct model.  In cases where the evidence strongly favors a particular model compared to the others, or where each model yields very similar posterior results, the choice is less difficult or of little consequence.

To handle this issue, we employ a Bayesian model averaging (BMA; \citealp{Hoeting99}) technique to obtain $P({L_d \mid \D})$, effectively marginalizing over all models that we consider.  To calculate this, we must first assume that one of the models considered is correct (but that we do not know which one is), and then choose a prior probability $P(\M_k)$ for each being the true model.  If there are $K$ models to consider, then the marginalized posterior is given by
\begin{equation}
P({L_d \mid \D}) = \sum_{k=1}^K P({L_d, \M_k \mid \D}), \label{eq:post_marg_over_M}
\end{equation}
where $P({L_d, \M_k \mid \D})$ is the joint posterior distribution describing the true value of $L_d$ and whether $\M_k$ is the correct model.  By the definition of joint probability we can rewrite this as 
\begin{equation}
P({L_d \mid \D}) = \sum_{k=1}^K P({L_d \mid \D, \M_k})P({\M_k \mid \D}), \label{eq:post_marg_over_M2}
\end{equation}
where $P({\M_k \mid \D})$ is the marginalized posterior probability for $\M_k$ being the true model, which is given by
\begin{equation} 
P({\M_k \mid \D}) = \frac{P({\D \mid \M_k})P(\M_k)}{\sum_{j=1}^K P({\D \mid \M_j})P(\M_j)}. \label{eq:marg_model_post}
\end{equation}
Here, $P({\D \mid \M_k})$ is the Bayesian evidence for model $\M_k$ given by Equation \eqref{eq:evidence}.

To apply BMA, we assume that all eight of our bad-measurement models have equal prior probability of being the correct one and hence choose $P(\M_k) = 1/K= 1/8$.  This assumption causes a number of cancellations in Equation \eqref{eq:marg_model_post}, and hence by plugging this into Equation \eqref{eq:post_marg_over_M2} the marginalized posterior for $L_d$ effectively becomes the evidence-weighted model average:
\begin{equation}
P({L_d \mid \D}) = \frac{\sum_{k=1}^8 P({L_d \mid \D, \M_k})P({\D \mid \M_k})}{\sum_{j=1}^8 P({\D \mid \M_j})}. \label{eq:post_marg_over_M2_final}
\end{equation}
Consequently, models with low Bayesian evidence for them will have little impact on our net posterior results, while models with comparable Bayesian evidence with each other will provide similar contributions to our net posterior results.  

\subsubsection{Choosing Priors for $L_d$ and $\Theta_k$} \label{sec:priors}
The Bayesian priors for the parameters included in our bad-measurement models (i.e., those contained in $\Theta_k$) are the same flat PDFs chosen in \S2.3.2 of \citet{Licquia1}.  As mentioned above, the dynamical range of scale lengths measured for massive spiral galaxies falls in the range of $1\lesssim L_d\lesssim10$ kpc.  Therefore, we have chosen a flat Bayesian prior where $P(L_d)$ = 1/9 where $1\leq L_d\leq10$ kpc, and zero otherwise.  This effectively allows our posterior results to be determined purely from the data.  For convenience, we tabulate the range that these flat priors span below in Table \ref{table:priors}.

\begin{deluxetable}{ccccc}
\tablenum{2}\label{table:priors}
\tablewidth{\columnwidth}
\tablecaption{Ranges for Flat Priors}
\tablehead{$L_d$ (kpc) & $f_\text{good}$ & $n$ & $Q$ & $F$}
\startdata
$[0,9]$ & $[0,1]$ & $[1,4]$ & $[0,1]$ & $[\sigma_i^\text{MIN}/\mu^\text{MED}, 1]$
\enddata
\tablecomments{The prior on $F$ is designed so that the free-$F$ model explores enforcing a floor value on error estimates ($=F\times\mu^\text{MED}$) ranging from the smallest value in the dataset ($\sigma_i^\text{MIN}$) up to the median central value ($\mu^\text{MED}$).  As reminder, we use a value of 2.5 kpc for $\mu^\text{MED}$ throughout this study; this is relevant for the free-$Q$ model as well, which explores adding extra uncertainty in quadrature to the nominal error estimates ($=Q\times\mu^\text{MED}$).}
\end{deluxetable}

\subsubsection{Corrections Toward a Uniform Dataset} \label{sec:corrections}
As described in \S\ref{sec:data_models}, the details and assumptions that go into models of the MW can vary considerably from study to study, which could be to blame for any inconsistencies amongst the $L_d$ estimates that we have collected.  While it is not possible to correct for all differences, it is important to attempt to place each of these estimates on equal footings, as much as possible, by normalizing them to a common set of assumptions.  This should optimize our meta-analysis by placing each measurement on a level playing field, and hence our posterior results should best reflect our true knowledge of $L_d$ given all of the observational data that measurements have utilized.

Unfortunately, the complexity of MW models and the correspondingly large number of free parameters that they generally include make this a difficult job.  For example, disk models that include a central truncation or ``hole'', which itself is parameterized by a scale radius, $L_{\text{hole}}$, tend to produce a strong anti-correlation between $L_d$ and $L_{\text{hole}}$, which can bias results P\&R04.  Similarly, including a cutoff radius, as opposed to extrapolating to $R=\infty$, may systematically reduce $L_d$.  Another dichotomy in Galactic models is that some incorporate substructure components (e.g., spiral arms and rings) and some do not.  These types of variations are practically impossible to correct for without redoing each analysis.  We note that the BMM meta-analysis should yield robust results so long as the plurality of measurements are constraining the same fundamental parameter.  If other models are sufficiently different in nature that they yield $L_d$ values that are systematically offset from the consensus, they will neither cause significant offsets in the posterior probability distribution nor decrease the breadth of that distribution in an unwarranted way (in contrast to the IVWM).

One assumption that can be corrected for is the authors' choice of $R_0$.  While variations in the adopted value of $R_0$ appear to have negligible impact on the $L_d$ results from star count studies (see, e.g., N95; M\&vA98), and hence the majority of our dataset, it does have a measurable effect on measurements based on integrated flux profiles.  For example, K91 explicitly state that their distance measurements will scale proportionately to $R_0$, while F98 published his results as a function of $R_0$ showing $L_d$ variations to be consistent with a linear scaling law, and D\&S01 quote their $L_d$ estimate in units of $R_0$.  Therefore, we assume that integrated flux measurements scale linearly when changing the choice of $R_0$ (cf. \citealp{Sackett97} and \citealp{Hammer}, who assume $L_d$ estimates from all measurement techniques scale proportionately with $R_0$).

Consistent with \citet{Licquia1}, we have adopted $R_0 = 8.33\pm0.35$ kpc from the work of \citet{Gillessen}, both because it is based upon a geometric technique robust to the systematic uncertainties in the distance ladder, and because it provides a thorough analysis of both statistical and systematic uncertainties.  This is a conservative choice as the error estimate is broad enough to be consistent with estimates from a variety of other measurement techniques, and hence describes well the current knowledge of this parameter.  We renormalize each $L_d$ estimate from integrated light measurements to reflect this choice of prior.  We do this using the ``strict-prior'' approach described in \citet{Licquia1}, where we force our posterior results to reflect our choice of prior by assuming that $P({R_0 \mid \D}) = P(R_0)$.  Hence, we assume that the observed variations amongst $L_d$ measurements contain no useful information about the true value of $R_0$.  This can be achieved through Monte Carlo simulations, which we describe in \S\ref{sec:MC} below.

Other assumptions that we make or alterations to particular estimates in Table 1 are as follows:
\begin{itemize}
\item In cases where an $L_d$ result is presented without a formal error estimate, we ascribe a 25\% error bar.
\item In almost all cases the scale length estimate describes the thin disk of the MW.  However, a few cases exist where a thick disk is also included and appears as the dominant component of the Galaxy.  Therefore, avoiding confusion due to nomenclature (e.g. ``thin'', ``thick'', ``old'', ``young'', etc.), we include the $L_d$ estimate associated with the most massive disk component included in the model.  This pertains primarily to the Galactic models in O\&L93 and L\&L00.  However, any studies that solely measure the scale length of stars which unambiguously belong to the thick disk are excluded \citep[e.g.][]{Robin96,Zheng01}
\item The estimate made by P98 relies on assuming that all stars observed to fall into the apparent magnitude bin of $9<K<10$ are of type K2--K5 III.  To account for the possibility that this assumption may be in error (especially given the minimal detail presented in this study and the lack of systematics checks) we conservatively augment the error estimate to 25\% of the central value.
\item The estimate made by M\&vA98 relies on applying a rather outdated model of extinction to photographic plate data taken from the mid-plane of the Galaxy.  Given this combination we conservatively augment the error estimate to 25\% of the central value.
\item The estimate made by C99 appears to correspond to the mean and approximate standard error of the $L_d$ values produced by fitting their model to four different fields of view.  Given that this neglects the contribution of uncertainties in their model assumptions as well as in the formal fitting of the data, we augment the error estimate to 25\% of the central value.
\item The estimate from D\&S01 comes from the mean and standard deviation of values presented in their Table 4 to incorporate uncertainties in the fixed parameters of the model.  We have calculated the standard deviation in the $L_d$ values from Table 5 and 6 in this paper to account for the uncertainty due to alternative stellar and spiral models, respectively.  Each of these values have been added in quadrature to the nominal error estimate to produce the error bar that we use in our analysis.
\end{itemize}

We note here that there are several estimates in the literature that we have assumed are superseded by an estimate we have included in our set, due to strong overlap in observational data, Galactic models employed, contributing authors (and hence assumptions), etc., and therefore have been excluded from this study.  These include \citet{Ruphy96} superseded by P\&R04, \citet{Spergel96} superseded by D\&S01, \citet{Larsen96} superseded by L\&H03, and \citet{Robin03} superseded by P\&R04.  We also exclude the estimates from \citet{Bovy12} and \citet{Cheng12} from our analyses as these rely critically on segregating stars into subpopulations based on their $\alpha$-element abundance ratios, a technique that is not viable for measuring extragalactic scale lengths, which we ultimately want to compare our results to.

With all of these assumptions in place, we present our renormalized dataset in Table \ref{table:HB_data}.  Here we have identified them as either visible or IR measurements, and show the error estimates updated to the assumptions listed above.  We also list the value of $R_0$ assumed by the author, and indicate the appropriate scaling relation by $\alpha$, where $L_d\propto R_0^\alpha$.  Error bars are scaled similarly to central values with $R_0$, such that logarithmic error bars remain constant.

\begin{deluxetable}{lcccc}
\tablenum{3}\label{table:HB_data}
\tablewidth{\columnwidth}
\tablecaption{Milky Way Radial Scale Length Measurements Renormalized For BMM Analysis}
\tablehead{Reference & $L_d\pm$1$\sigma$ & $R_0$ & $\alpha$ & Spectral \\
 & (kpc) & (kpc) & & Regime}
\startdata
K91 & $3.00\pm0.54$ & 8.00 & 1 & Infrared \\
RM91 & $2.50\pm0.63$ & 8.75 & 1 & Infrared \\
Y\&Y92 & $3.80\pm0.27$ & 8.00 & 0 & Visible \\
R92 & $2.50\pm0.30$ & 8.50 & 0 & Visible \\
O\&L93 & $2.60\pm0.50$ & 7.90 & 0 & Infrared \\
N95 & $4.50\pm1.13$ & 8.06 & 0 & Visible \\
O96 & $2.30\pm0.60$ & 8.09 & 0 & Visible \\
P98 & $2.10\pm0.53$ & 8.50 & 0 & Infrared \\
F98 & $2.60\pm0.65$ & 8.50 & 1 & Infrared \\
B98 & $4.01\pm1.00$ & 8.60 & 0 & Visible \\
M\&vA98 & $6.00\pm1.50$ & 8.50 & 0 & Visible \\
C99 & $2.25\pm0.56$ & 8.50 & 0 & Visible \\
L\&L00 & $2.30\pm0.58$ & 8.50 & 1 & Infrared \\
D\&S01 & $2.26\pm0.22$ & 8.00 & 1 & Infrared \\
O01 & $2.80\pm0.30$ & 8.00 & 0 & Infrared \\
S02 & $2.25\pm0.56$ & 8.00 & 0 & Visible \\
LC02 & $3.30\pm0.50$ & 7.90 & 0 & Infrared \\
L\&H03 & $3.50\pm0.30$ & 8.00 & 0 & Visible \\
P\&R04 & $2.35\pm0.67$ & 8.50 & 0 & Infrared \\
G05 & $2.80\pm0.25$ & 8.50 & 0 & Visible \\
B05 & $3.90\pm0.60$ & 8.50 & 0 & Infrared \\
B06 & $1.90\pm0.48$ & 8.60 & 0 & Visible \\
K07 & $2.25\pm0.56$ & 8.00 & 0 & Visible \\
J08 & $2.60\pm0.52$ & 8.00 & 0 & Visible \\
C11 & $3.70\pm1.00$ & 8.00 & 0 & Infrared \\
R12 & $2.20\pm0.55$ & 8.00 & 0 & Infrared \\
P13 & $2.12\pm0.20$ & 8.00 & 0 & Infrared \\
LC\&M14 & $2.00\pm0.40$ & 8.00 & 0 & Visible \\
M15 & $2.34\pm0.48$ & 8.00 & 0 & Visible
\enddata
\tablecomments{See the Table 1 footnotes for reference abbreviations.  In cases where the authors do not specify the $R_0$ assumed, we assume they have used the IAU standard value of 8.5 kpc \citep{IAU}.  In column 4, $\alpha$ reflects the assumed scaling law, $L_d\propto R_0^\alpha$, based on the measurement technique.}
\end{deluxetable}

\subsubsection{Monte Carlo Techniques} \label{sec:MC}
Combining all of the assumptions and techniques described above requires us to perform a set of Monte Carlo simulations.  For convenience, we now summarize the full process of obtaining our results.  We begin by drawing 10$^3$ realizations of $R_{0,i}$ from $P(R_0)=8.33\pm0.35$ kpc, which is the result from \citet{Gillessen} discussed in the previous section.  For each $R_{0,i}$ we first renormalize each independent estimate of $L_d$ in Table \ref{table:HB_data} from the $R_0$ assumed by the author to $R_{0,i}$ according to the appropriate scaling law, yielding \D.  Next, for each bad-measurement model $\M_k$, we calculate the likelihood, $P({\D \mid L_d, \Theta_k, \M_k, R_{0,i}})$, from Equations (2)--(5) in \citet{Licquia1} and determine the Bayesian evidence using Equation \eqref{eq:evidence} above.

Following the strict-prior methodology from \citet[][i.e., assuming $P({R_0 \mid \D}) = P(R_0)$]{Licquia1}, we next calculate the marginalized posterior for the MW's disk scale length by
\begin{align}
& P({L_d \mid \D, \M_k}) = \nonumber \\
&\quad \frac{1}{1000} \sum_{i=1}^{1000} \int \frac{P({\D \mid L_d, \Theta_k, \M_k, R_{0,i}})P(L_d)P(\Theta_k)}{P({\D \mid \M_k, R_{0,i}})} \,\text{d}\Theta_k.
\end{align}
Lastly, we implement Bayesian model averaging (BMA) by calculating the average posterior of the models weighted by the \emph{mean} evidence by averaging the results from the 10$^3$ realizations of $R_{0,i}$:
\begin{multline}
P({L_d \mid \D}) = \\ \frac{\sum_{k=1}^8 P({L_d \mid \D, \M_k}) \sum_{i=1}^{1000} P({\D \mid \M_k, R_{0,i}})}{\sum_{j=1}^8 \sum_{i=1}^{1000} P({\D \mid \M_j, R_{0,i}})}.
\end{multline}
We find that all of our results are entirely unchanged if we use the median evidence values instead of the mean.

For ease of comparison amongst our models, we will recast our presentation of the Bayesian evidence in terms of the \emph{Bayes Factor}, which in our framework is calculated by
\begin{equation}
\B_k = \frac{P({\D \mid \M_k})}{P({\D \mid \M_{\text{all-good}}})} = \frac{\sum_{i=1}^{1000} P({\D \mid \M_k, R_{0,i}})}{\sum_{j=1}^{1000} P({\D \mid \M_{\text{all-good}}, R_{0,j}})}. \label{eq:Bayes_factor}
\end{equation}
This represents the ratio of the posterior odds to the prior odds in favor of model $\M_k$ over the all-good model, our null hypothesis model.  The standard rule-of-thumb \citep[][and references therein]{KassRaftery95} applied in the context of this study means that finding $\log \B_k>\sim2$ indicates a statistically significant favorability for model $\M_k$ over the all-good model.  Values of $\log \B_k$ in the range of $\sim$0.5--1 and $\sim$1--2 are considered as ``substantial'' and ``strong'' favorability, respectively, but are not necessarily large enough to decisively deem $\M_k$ as the superior model, and negative values of course favor the all-good model over $\M_k$ on the same scale.  One can also conveniently reframe this type of comparison between any two models $\M_k$ and $\M_j$ by simply calculating $\log \B_k - \log \B_j$.

We note here that we have also tested for the effect of treating $R_0$ as a free parameter in our modeling of the data and adopting the same prior $P(R_0)$.  This breaks our assumption that $L_d$ estimates contain no useful information about $R_0$, and eliminates the need for marginalization through the Monte Carlo framework laid out above.  We find that the marginal posterior results for $L_d$ are nearly identical to those of our standard treatment, which is likely due to only a small subset of the $L_d$ dataset being dependent on $R_0$.  Nevertheless, we believe that implementing the strict-prior methodology is well motivated and worth the added complexity for our purposes.

\section{Results} \label{sec:results}

In Table \ref{table:HB_results} we present the results for each of the eight bad-measurement models.  This includes the number of free parameters, $N_{\text{free}}$, the median and 68\% credible interval measured from the marginalized posterior $P({L_d \mid \D, \M_k})$, and the corresponding $\log\B_k$ value.  We have broken our results down into three scenarios: analyzing the entire dataset provided in Table \ref{table:HB_data}, analyzing only those measurements made in the visible, and analyzing only those measurements made in the IR.  Lastly, for each of these three scenarios we list the median and 68\% credible interval measured from the posterior marginalized over all models, $P({L_d \mid \D})$, calculated via BMA (see \S\ref{sec:BMA}), which serve as our nominal results.  To easily compare our results with a standard combination of the data, we include in Table \ref{table:HB_results} the inverse variance-weighted mean (IVWM) of the $L_d$ estimates. We note that the result from the all-good model is equivalent to the IVWM of the estimates after scaling them to reflect assuming $R_0=8.33$ kpc according to the appropriate relation indicated in Table \ref{table:HB_data}.

\begin{deluxetable}{lccc}
\tablenum{4}\label{table:HB_results}
\tablewidth{\columnwidth}
\tablecaption{BMM Results for the Milky Way Radial Scale Length}
\tablehead{Model ($\M_k$) & $N_{\text{free}}$ & $L_d\pm$1$\sigma$ & $\log\B_k$ \\
 & & (kpc) & }
\startdata
\multicolumn{4}{c}{\underline{IR estimates only}} \\[1ex]
free-$n$ & 3 & $2.48^{+0.14}_{-0.13}$ & $-0.28$ \\
free-$Q$ & 3 & $2.51^{+0.16}_{-0.14}$ & $-0.27$ \\
free-$F$ & 3 & $2.53^{+0.17}_{-0.15}$ & $-0.24$ \\
all-bad free-$n$ & 2 & $2.48\pm0.14$ & $-0.66$ \\
all-bad free-$Q$ & 2 & $2.56^{+0.17}_{-0.15}$ & $-0.47$ \\
all-bad free-$F$ & 2 & $2.62^{+0.16}_{-0.15}$ & $-0.37$ \\
$P_\text{bad}$-flat & 2 & $2.47\pm0.12$ & $-0.93$ \\
all-good & 1 & $2.48\pm0.11$ & ----- \\
\\[-1.5ex]
\multicolumn{2}{l}{BMA result} & $2.51^{+0.15}_{-0.13}$ & \\
\multicolumn{2}{l}{IVWM} & $2.46\pm0.11$ & \\
%\multicolumn{2}{l}{IVWM ($R_0$=8.33 kpc)}& $2.48\pm0.11$ &  \\
\hline
\\[-1.5ex]
\multicolumn{4}{c}{\underline{Visible estimates only}} \\[1ex]
free-$n$ & 3 & $2.68^{+0.21}_{-0.19}$ & $1.94$ \\
free-$Q$ & 3 & $2.64\pm0.19$ & $1.97$ \\
free-$F$ & 3 & $2.64^{+0.18}_{-0.17}$ & $1.94$ \\
all-bad free-$n$ & 2 & $2.85\pm0.20$ & $1.97$ \\
all-bad free-$Q$ & 2 & $2.77\pm0.21$ & $1.95$ \\
all-bad free-$F$ & 2 & $2.70^{+0.21}_{-0.18}$ & $1.97$ \\
$P_\text{bad}$-flat & 2 & $2.52^{+0.17}_{-0.16}$ & $1.03$ \\
all-good & 1 & $2.85\pm0.11$ & ----- \\
\\[-1.5ex]
\multicolumn{2}{l}{BMA result} & $2.71^{+0.22}_{-0.20}$ & \\
\multicolumn{2}{l}{IVWM} & $2.85\pm0.11$ & \\
\hline
\\[-1.5ex]
\multicolumn{4}{c}{\underline{Visible and IR estimates combined}} \\[1ex]
free-$n$ & 3 & $2.57\pm0.12$ & $2.39$ \\
free-$Q$ & 3 & $2.60^{+0.14}_{-0.13}$ & $2.69$ \\
free-$F$ & 3 & $2.62\pm0.12$ & $2.85$ \\
all-bad free-$n$ & 2 & $2.66\pm0.12$ & $1.88$ \\
all-bad free-$Q$ & 2 & $2.69^{+0.13}_{-0.12}$ & $2.64$ \\
all-bad free-$F$ & 2 & $2.66\pm0.12$ & $3.00$ \\
$P_\text{bad}$-flat & 2 & $2.50\pm0.10$ & $1.26$ \\
all-good & 1 & $2.66\pm0.08$ & ----- \\
\\[-1.5ex]
\multicolumn{2}{l}{BMA result} & $2.64\pm0.13$ & \\
\multicolumn{2}{l}{IVWM} & $2.65\pm0.08$ & \\
%\multicolumn{2}{l}{IVWM ($R_0$=8.33 kpc)} & $2.66\pm0.08$ &  \\
\enddata
\tablecomments{For each dataset, the first eight rows shows the results from each of our bad-measurement models detailed in \citet{Licquia1}.  $N_{\text{free}}$ denotes the number of free parameters that each model contains.  Below this point, we show the result of applying Bayesian model averaging (BMA) to all eight models, which in this study reflects the median and 68\% credible interval measured from the Bayesian evidence-weighted average of all $P({L_d \mid \D, \M_k})$, since we assume as a prior that each model is equally probable (see \S\ref{sec:BMA}).  Finally, for comparison, we also list the inverse variance-weighted mean (IVWM) from using the nominal central values and error estimates in Table \ref{table:HB_data}.  Note that the result from the all-good model is equivalent to calculating the IVWM \emph{after} scaling individual estimates to reflect our prior on $R_0$ where applicable.}
\end{deluxetable}

\subsection{Parameters Describing the Consistency of the Data} \label{sec:params_data}
\begin{figure*}[ht]
%\begin{figure}[ht]
\centering
\includegraphics[width=\textwidth, trim=.1in 0in 0in 0in, clip=true]{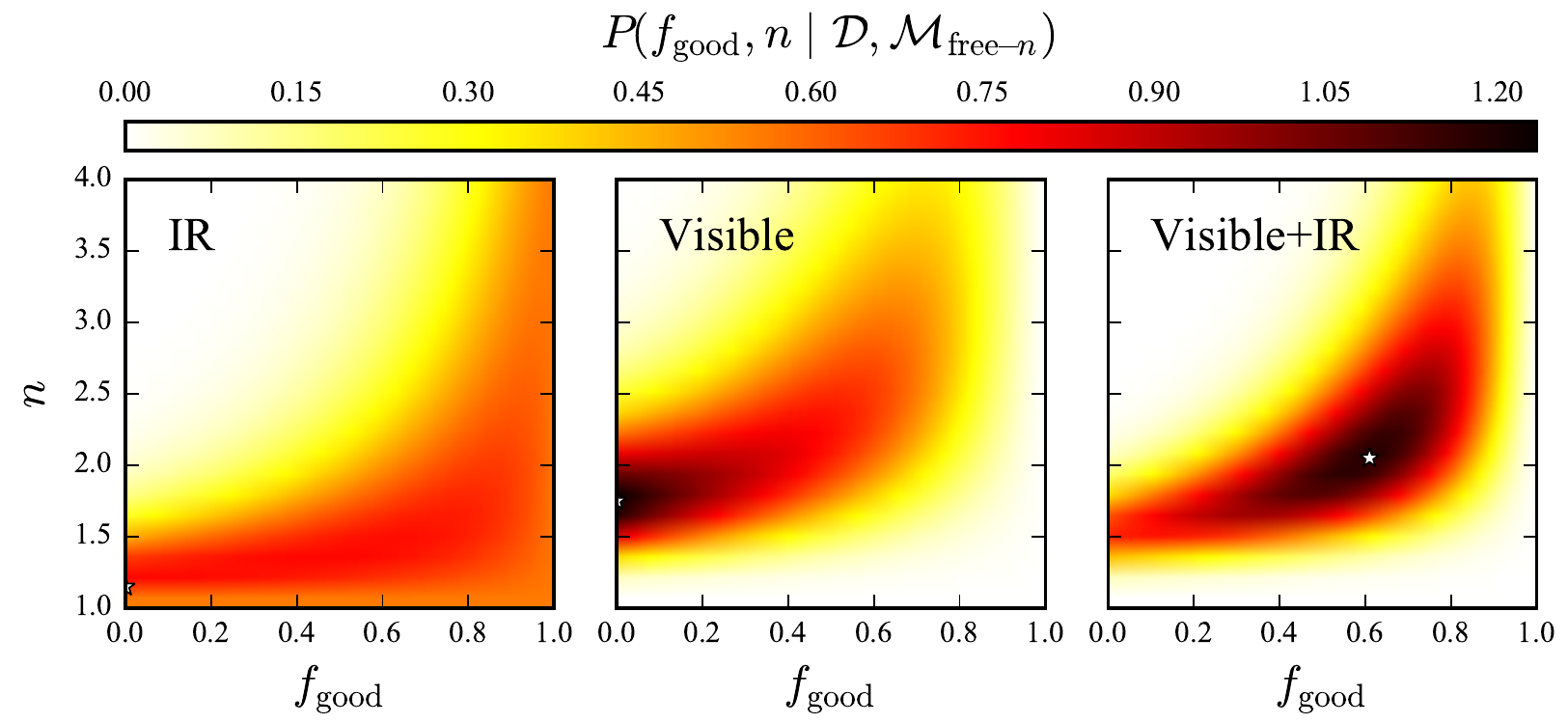}
\caption{Quantifying the overall fidelity of Milky Way disk scale length measurements, \D, in terms of both the frequency of accurate estimates, $f_\text{good}$, and the error bar correction factor for inaccurate measurements, $n$.  Each panel shows the joint posterior distribution, $P({f_\text{good}, n \mid \D, \M_{\text{free-}n}})$, yielded from the free-$n$ model, $\M_{\text{free-}n}$, when analyzing $L_d$ estimates measured in the infrared (IR; left panel) or the visible (middle panel), and when analyzing both the visible and IR data together (right panel).  We have marked the peak probability location in each panel by a white star.  The set of IR estimates alone are in good overall consistency with one another, as the posterior peaks at $n\sim1$, the limit where the correction for bad estimates goes away (note that values of $n=1$ or $f_\text{good}=1$ reduce to the all-good model).  In contrast, the visible estimates alone, due to the strong tension amongst them, appear to contain a significant number that are biased or erroneous.  Here, the posterior peaks at $f_\text{good}=0$ and $n=1.75$, corresponding to the case where all measurements are treated as having error estimates that are 75\% too small.  The right panel indicates that there is good consistency amongst some of the visible and IR data together, as it yields an intermediate result between analyzing the visible and IR data separately.}
\label{fig:joint_posts}
%\end{figure}
\end{figure*}

As indicated by the Bayes factors listed in Table \ref{table:HB_results}, the IR data are all consistent enough that the all-good model yields a satisfactory fit with no additional free parameters, causing it to be favored by the evidence calculations.  This corresponds to assuming high fidelity for all estimates.  However, when the visible measurements are included in the dataset, either by analyzing them alone or in combination with the IR measurements, there appears to be enough tension amongst them that the all-good model becomes strongly disfavored.  To illustrate this further, we show in Figure \ref{fig:joint_posts} the results for parameters describing the consistency of the data from the free-$n$ model.  Each panel shows the joint posterior, $P({f_\text{good},n \mid \D, \M_{\text{free-}n}})$, for the fraction of ``good'' (i.e., accurate) estimates in the set, $f_\text{good}$, and the correction factor needed for error estimates from bad measurements, $n$.  The white star marks the peak of the distribution.

The left panel shows the result from analyzing the IR estimates alone.  Here, we find a relatively broad posterior that peaks at $n=1.15$ and $f_\text{good}=0.03$.  That is to say, the likelihood function for the free-$n$ model is maximized under the assumption that nearly all measurements of $L_d$ in the IR have underestimated errors by 15\%.  Note, however, that this is also to say that the posterior peaks nearly along the $n=1$ slice through this plane, where the free-$n$ model (with 3 free parameters) becomes equivalent to the all-good model (with 1 free parameter).  This explains why the latter is favored by the Bayesian evidence --- it provides a similar description of the data without requiring superfluous parameter space.

The middle panel shows the result from analyzing only the visible $L_d$ estimates, which appear to be in significant tension with one another.  In this case, the posterior is more strongly peaked and is maximized at $n=1.75$ and $f_\text{good}=0$, where error estimates on all measurements are increased by 75\%.  Note that this is where the free-$n$ model becomes equivalent to the all-bad free-$n$ model, though each yields very similar $\log\B_k$ values.  This represents an example where the added parameter space yields a large enough improvement in likelihood values that they roughly balance.

Finally, the right panel shows the joint posterior from analyzing all (visible+IR) measurements in Table \ref{table:HB_data}, which appears to be an intermediate result compared to those from treating the visible and IR data separately.  The posterior is maximized when assuming that $\sim$40\% of the set have overly-optimistic error estimates by a factor of $\sim$2.  This indicates that there is a good level of consistency between many of the 14 IR and 15 visible measurements, but that there are some estimates that are in strong tension with the rest.  Note the covariance between $f_\text{good}$ and $n$ in the model, which shows that there are multiple ways to account for this tension.  For example, upon further inspection of Table \ref{table:HB_data}, it seems there are a few strongly-peaked (i.e., with small estimated errors), similarly valued estimates (Y\&Y92 and L\&H03) that are in modest tension with a number of lower-valued, broader estimates (O96, S02, B06, K07, J08, LC\&M14, M15); treating the former as problematic would result in larger values for both $f_\text{good}$ and $n$, whereas considering the latter set as problematic would correspond to lower values of $f_\text{good}$ and $n$ (see Fig. \ref{fig:Ld_posts} for a visual comparison, which we discuss below).  Likely there are several other ways to reduce the tension amongst these measurements; each would lead to a broader posterior distribution than the IVWM by incorporating this extra uncertainty.  This would typically be unaccounted for by other meta-analysis techniques.  We note that despite the wide varieties of models considered, our final BMA results are consistent with all of them at $<$1$\sigma$; it therefore appears unlikely that entertaining more models of bad measurements would significantly change our final results.

\begin{figure*}[ht]
%\begin{figure}[ht]
\centering
\includegraphics[width=\textwidth, trim=.1in .1in .1in .1in, clip=true]{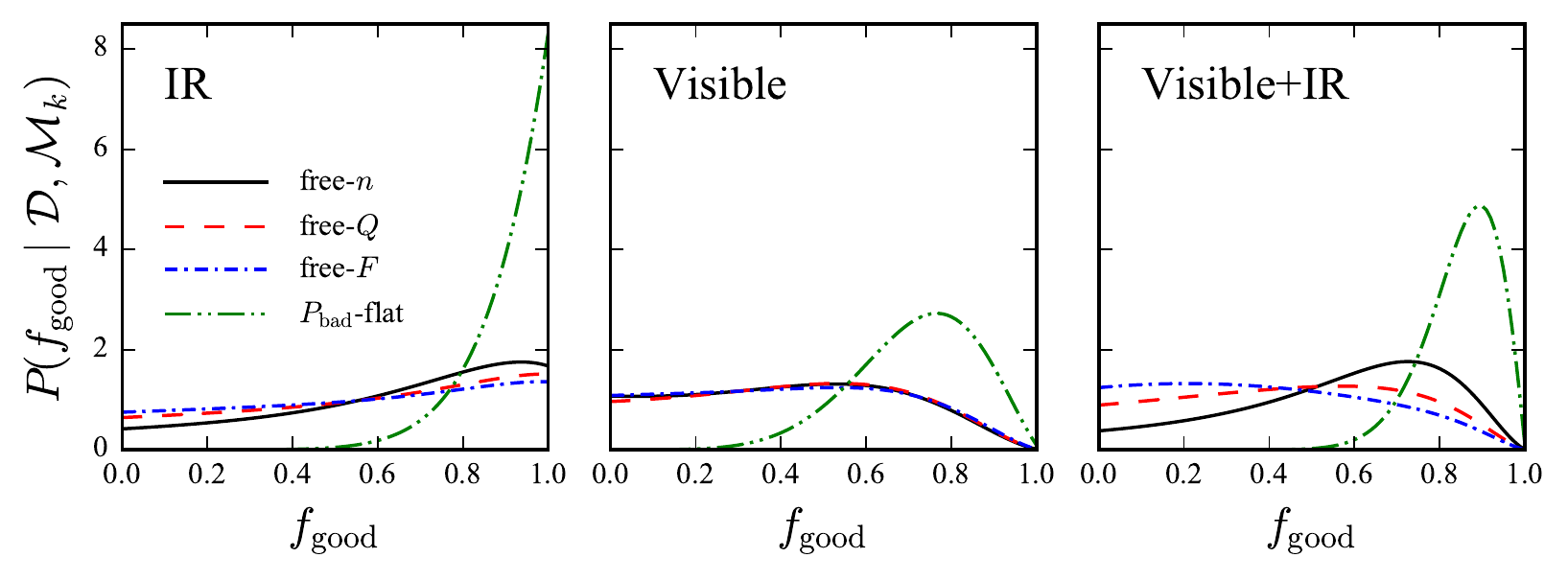}
\caption{Quantifying the frequency of accurate estimates, $f_\text{good}$, amongst our $L_d$ dataset, \D, using a variety of bad-measurement models, $\M_k$.  Each panel shows the marginal posterior for $f_\text{good}$, $P({f_\text{good} \mid \D, \M_k})$, from the four models listed in the left panel (see \S\ref{sec:hb} for a short description of these) after integrating over all other free parameters.  This is divided into cases of analyzing the IR estimates alone (left panel), the visible estimates alone (middle panel), and the visible and IR estimates together (right panel).  As similarly described in Fig. \ref{fig:joint_posts}, the IR data alone is in good overall consistency, leading to optimal values of $f_\text{good}=1$.  The visible data alone, however, appear to be contaminated by erroneous measurements as the posteriors from all models peak at values of $f_\text{good}$ well below unity.  The results from analyzing the visible and IR data together are more divergent, with different models favoring differing values of $f_\text{good}$; i.e., some models are able to ease inter-measurement tension by altering only a few $L_d$ estimates, whereas others require expanding error estimates on a larger number of measurements.  In all scenarios, the $P_\text{bad}$-flat model favors higher values of $f_\text{good}$ than the other models; this is because it effectively discards a measurement when assuming it's inaccurate, which is a much more expeditious route to easing inter-measurement tension.  As is evident in Table \ref{table:HB_results}, the associated loss of information from this approach is more detrimental to the Bayesian evidence though, preventing the $P_\text{bad}$-flat model from being the most favorable (see the discussion of Bayes Factors in \S\ref{sec:MC}).  Our final results for $L_d$ are produced by averaging over the possibility of each model being correct (see \S\ref{sec:BMA}).}
\label{fig:fgood_posts}
%\end{figure}
\end{figure*}

To provide a picture of how different models compare in identifying bad measurements, Figure \ref{fig:fgood_posts} displays the marginal posterior for $f_\text{good}$ for all four of those that allow for the possibility that \emph{some} of the measurements are problematic (in all others $f_\text{good}$ has been either set to zero --- the entries labeled as ``all-bad'' in Table \ref{table:HB_results} --- or set to unity in the case of the ``all-good'' entries).  Consistent with the story from Figure \ref{fig:joint_posts}, we find that $P({f_\text{good} \mid \D, \M_k})$ peaks at $f_\text{good}=1$ for all models when analyzing the IR estimates alone, but peaks well below $f_\text{good}=1$ when analyzing the visible estimates alone.  Interestingly, these curves show more diversity when analyzing the visible and IR estimates in combination and peak at a range of values of $f_\text{good}$.

In all cases, the $P_\text{bad}$-flat model produces a tighter posterior distribution that supports larger values of $f_\text{good}$, as it is effectively able to discard measurements entirely when treating them as bad.  However, the corresponding loss of information weakens the overall likelihood function and hence yields a smaller $\B_k$, preventing this from being the prevailing model.  In contrast, the three other models, each uniquely remedying the effect of bad measurements instead of discarding them, give much more probability to lower values of $f_\text{good}$, while also yielding larger $\B_k$.  In other words, the Bayesian evidence in this application more strongly supports models that attempt to correct inaccurate error estimates for measurements of the Galactic scale length, rather than assuming that outlying measurements are too problematic to contain any useful information.  This is the same message in the panel corresponding to the IR data alone, except that for all models the marginalized posterior peaks at $f_\text{good}=1$.  As a result, each model looks quite similar to the all-good model (where $f_\text{good}$ has been set to unity); it is reasonable then that the Bayesian evidence deems the all-good model superior, as it does not contain unneeded free parameters.

\subsection{Marginalized Posterior Results for $L_d$} \label{sec:post_results_ld}
From the values in Table \ref{table:HB_results}, one can find several instances where two models are supported by similar values of $\B_k$, but have non-negligible differences in their $P({L_d \mid \D, \M_k})$.  This is especially common amongst our results from analyzing the visible estimates alone; here, a variety of bad-measurement models yield $\log\B_k$ values of $\sim$2 (indicating strong favorability over the all-good model), yet central values for $L_d$ that span from 2.64--2.85 kpc, a range that is comparable in size to the 1$\sigma$ error estimate from any single model.  This provides a cautionary tale: two models may be quite comparable in how well they can describe the data, but still lead to somewhat differing results (in this case, all models are consistent with each other at the $\sim$0.8$\sigma$ level or better).

If we did not employ BMA, we would be left in the position of having to choose a winning model for more arbitrary reasons.  In our example, one could choose the free-$Q$ model as it maximizes the evidence (though not uniquely), or one could choose the all-bad free-$Q$ model (where $f_\text{good}$ has been set to zero, instead of treated as a free parameter) in an effort to be conservative as it yields the largest error estimate and a central value towards the center of the pack.  This illustrates a new source of uncertainty -- i.e., model selection uncertainty -- that needs to be accounted for.  Hence, we have implemented the Bayesian model averaging (BMA) technique, detailed in \S\ref{sec:BMA}, to produce the final posterior distribution for $L_d$; the method effectively marginalizes over the possibility of all the models we have considered being the correct model.

\begin{figure*}[ht]
%\begin{figure}[ht]
\centering
\includegraphics[width=\textwidth, trim=.1in .1in .1in .1in, clip=true]{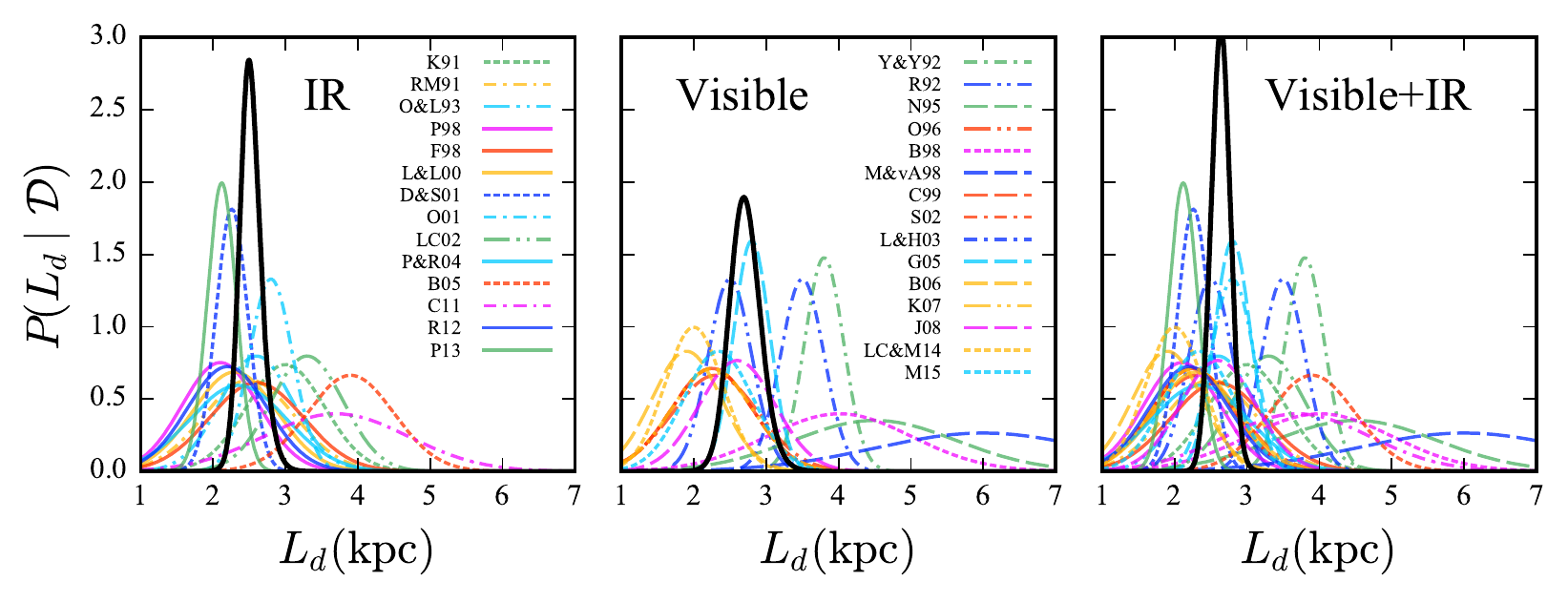}
\caption{Marginal posterior results, $P({L_d \mid \D})$, for the exponential scale length, $L_d$, of the Milky Way's disk yielded from BMM meta-analysis of IR measurements (left panel), visible measurements (middle panel), and visible and IR measurements combined (right panel), shown by a thick black curve.  Each is produced via a Bayesian model averaging (BMA) technique described in \S\ref{sec:BMA}, where we have effectively marginalized over the probability of each bad-measurement model we have considered as being correct, and hence incorporates all model selection uncertainties.  To provide a visual comparison, we have overlaid the individual estimates of $L_d$ that have gone into our meta-analysis as colored textured curves.  The abbreviations for each study can be found in the notes for Table 1.}
\label{fig:Ld_posts}
%\end{figure}
\end{figure*}

In Figure \ref{fig:Ld_posts} we show the finalized posterior results for the Galactic scale length from BMA for the IR, visible, and visible+IR datasets, which correspond to $2.51^{+0.15}_{-0.13}$ kpc, $2.71^{+0.22}_{-0.20}$ kpc, and $2.64\pm0.13$ kpc, respectively.  So that one may visually compare the individual estimates to the aggregate result that they have been incorporated into (the thick black curves), we have overlaid the probability distribution corresponding to each estimate in colored, dashed/dotted curves.  In total, each result represents our best estimate of the Galactic scale length in the IR, visible, and visible+IR regimes given the array of values available in the literature.

\subsection{Impact of $R_0$ Prior} \label{sec:r0dependence}
As discussed above, a small subset of the $L_d$ estimates we have collected, particularly those measured from maps of integrated IR starlight, scale proportionately with the choice of $R_0$ (see Table \ref{table:HB_data}).  As a reminder, we have adopted a prior of $R_0 = 8.33\pm0.35$ kpc and have utilized Monte Carlo techniques (see \S\ref{sec:MC}) to propagate this into our posterior results for $L_d$.  Hence, our IR and visible+IR results, too, should be dependent on the choice of $R_0$ to some degree.  We will now discuss how our prior for $R_0$ affects the results we have found for $L_d$, which will allow one to update them as our knowledge of $R_0$ becomes more accurate.

We have first tested to see how much the adopted uncertainties in $R_0$ contribute to the overall uncertainties in $L_d$.  To do so, in place of sampling from our $R_0$ prior $10^3$ times, we have repeated our analysis using only a single iteration of scaling all measurements to $R_0 = 8.33$ kpc, thereby assuming 100\% accuracy in our knowledge of this parameter.  We find that this yields identical posterior results, indicating that $R_0$ uncertainties are negligible in the overall error budget for $L_d$.

Next, we have investigated how our results change if we adopt $R_0$ values anywhere in the range of 7.5--9 kpc, which is closely equal to the 2$\sigma$ range from by our nominal prior.  In line with our error analysis, this reveals only weak correlations between $R_0$ and $L_d$: the IR estimate is well described by $L_d = 0.12(R_0 - 8.33\,\text{kpc})+2.51^{+0.15}_{-0.13}$ kpc, whereas the visible+IR estimate is well described by $L_d = 0.06(R_0 - 8.33\,\text{kpc})+2.64\pm0.13$ kpc.  As a reminder, none of the visible $L_d$ estimates we have collected vary with the choice of $R_0$, and hence neither does the visible $L_d$ result found here.

\subsection{Tests of Alternative Assumptions and Robustness} \label{sec:robust}
As indicated in Table 1, many of the visible $L_d$ estimates we have included in our analyses pre-date the dust maps produced by SFD98, which provided a largely improved tool for modeling Galactic reddening and extinction.  We have tested for the effect of deweighting visible $L_d$ estimates that do not utilize the SFD98 dust maps by adding an extra 10\% error in quadrature to their nominal error bars and repeating our entire analysis.  Interestingly, in this scenario we find that when analyzing either the entire visible+IR dataset or just the visible estimates alone, all bad-measurement models yield $\log\B_k$ values of $\sim$0 or below; i.e., the data become much more consistent overall, and the all-good model is now the most favored.  After applying BMA, we find that our aggregate visible+IR scale length estimate becomes $2.58^{+0.11}_{-0.10}$ kpc and the aggregate scale length from visible estimates becomes $2.65\pm0.16$ kpc.  These values are 0.06 kpc lower than our nominal results (see Table \ref{table:HB_results}), but consistent with them at the 0.35$\sigma$ and 0.23$\sigma$ levels, respectively.

Presumably IR measurements are less impacted by dust effects so we have not performed a similar deweighting test for them.  However, a few of the IR estimates in Table \ref{table:HB_data} have relatively small error estimates ($\sim$0.2 kpc).  To test whether our posterior results are dominated by estimates with the smallest errors, we have reanalyzed the IR data alone after replacing the P13 error estimate of 0.20 kpc with the median error estimate of 0.55 kpc.  This serves to systematically decrease the overall tension among the estimates in our dataset, and, as expected, the favorability of the all-good model increases; i.e., $\log\B_k$ values for all other models become more negative.  On the other hand, the posterior result from each model, $P({L_d \mid \D, \M_k})$, shifts toward higher values of $L_d$, and after applying BMA the posterior becomes $P({L_d \mid \D}) = 2.60\pm0.14$ kpc.  This is consistent with our nominal result at the 0.4$\sigma$ level, and provides a good indication that our analysis method is robust.

Lastly, as noted above, other studies that have reviewed or analyzed Galactic $L_d$ estimates in the literature \citep[e.g.,][]{Sackett97,Hammer} have assumed that each is directly proportional to the value of $R_0$ adopted by each author.  This assumption does not appear to be clearly correct for star count analyses, which constitute the majority of our dataset.  In fact, several of these studies indicated that their results for $L_d$ are highly insensitive to the choice of $R_0$, whereas studies modeling the integrated light profile of the Galaxy almost unanimously noted a linear dependence on $R_0$.  This is likely because star count measurements are so local that geometrical effects that depend on $R_0$ prove negligible.  However, in case this is an error on our part, we have investigated how our results change if we assume that $L_d \propto R_0$ for all estimates uniformly.  The effects on values listed in Table \ref{table:HB_results} are as follows.  

When analyzing the visible and IR estimates together, the $\log\B_k$ values for each model increased by $\sim$1, indicating a substantial increase in tension amongst the data, and hence increased favorability for bad-measurement mixture models.  The 68\% credible intervals measured from $P({L_d \mid \D, \M_k})$ increased by $\sim$30--60\%, whereas the median values shifted by $<2\%$.  The net result from performing BMA becomes $P({L_d \mid \D}) = 2.63^{+0.18}_{-0.17}$ kpc (as compared to $2.64\pm0.13$ kpc for our preferred analysis).

When analyzing the visible estimates alone, the $\log\B_k$ values similarly increase by $\sim$1.  Changes in $P({L_d \mid \D, \M_k})$ were very similar to those described above for the visible+IR dataset.  The net result from performing BMA becomes $P({L_d \mid \D}) = 2.70^{+0.27}_{-0.23}$ kpc (as compared to $2.71^{+0.22}_{-0.20}$ kpc for our preferred analysis).

When analyzing the IR estimates only, $\log\B_k$ values decreased (became more negative) by $<0.2$ adding slightly more favorability to the all-good (non-mixture) model.  Median values and 68\% credible intervals measured from $P({L_d \mid \D, \M_k})$ increased by $\sim$1\% and $\sim$10\%, respectively, except for that the 1$\sigma$ ranges for the $P_\text{bad}$-flat and all-good models expanded by 25\% and 35\%, respectively.  The result from performing BMA becomes $P({L_d \mid \D}) = 2.53^{+0.17}_{-0.16}$ kpc (as compared to $2.51^{+0.15}_{-0.13}$ kpc for our preferred analysis).

By comparing to the values in Table \ref{table:HB_results}, the overall impact on our nominal results is minimal, with shifts in the central value far below the 1$\sigma$ level and error estimates increasing by $<\sim$30\%.  This may stem from the fact that the $R_0$ values assumed by these studies have a mean and standard deviation of 8.22 kpc and 0.27 kpc, comparable to the $R_0$ prior we have chosen.  Regardless, our results appear robust to varying assumptions about how star count-based estimates of $L_d$ scale with $R_0$.

\section{Discussion} \label{sec:discussion}
\subsection{Comparisons to Dynamical Estimates} \label{sec:comp_dynamical}
In this study we have exclusively analyzed estimates of $L_d$ that have been produced from photometric models of the MW.  We therefore have excluded many estimates that have been produced from dynamical mass models of the MW that are predominantly constrained by kinematic data.  We now provide a quantitative comparison of our results to these dynamical estimates.  

We first emphasize that, just like photometric estimates, dynamical estimates of $L_d$ too rest upon many modeling assumptions, and hence are prone to systematic errors.  These predominantly rely on fitting models of both the stellar and dark matter components of the MW to measurements of stellar or gaseous kinematics.  \citet{Schoenrich10} points out that those studies utilizing the asymmetric drift (AD) relation (e.g., \citealp{DehnenBinney98} and subsequent studies) are particularly vulnerable to systematics stemming from the metallicity gradient of the Galactic disk, and can violate the assumptions that underpin the AD technique when binning stars by color.  Studies that aim to model the Galactic rotation curve also generally suffer from a strong disk-halo degeneracy, and hence the $L_d$ estimates that they yield can be sensitive to choices regarding the disk mass-to-light ratio and dark matter halo parameters.  Furthermore, one should keep in mind that the degree to which dynamical mass traces the distribution of starlight is not well understood, and hence to some extent these comparisons may not be apples-to-apples.

It appears that in many (but not all) cases, dynamical estimates of $L_d$ are proportional to the adopted value of $R_0$.  In the following, if a particular estimate is clearly stated by the authors to depend linearly on $R_0$ then we have scaled this estimate to reflect the prior on $R_0$ chosen in this study ($8.33\pm0.35$ kpc) to make a better comparison, and we have marked these values with a dagger.  However, in all other cases the authors have made no indication as to the dependence of their $L_d$ result on the choice of $R_0$, have fit for both quantities simultaneously, or have stated that these quantities are independent, and so for them we compare to the nominal values they have quoted.

\renewcommand{\thefootnote}{$\dagger$}
There are numerous dynamical estimates for $L_d$ in the literature that compare well with our results, either being consistent at or below the $\sim$1$\sigma$ level or falling into the 1$\sigma$ ranges presented herein if lacking an error estimate.  Such studies include 
\citet{FuxMartinet94}, who combined literature estimates of several properties of the Galactic disk with an approximated AD Jeans equation to find $L_d=2.45\pm0.50$\footnote{Denotes where the authors have stated that their $L_d$ result scales linearly with the choice of $R_0$, and so we have renormalized it to reflect our prior of $R_0=8.33\pm0.35$ kpc.} kpc; 
\citet{Durand96}, who modeled the Galactic potential using the radial velocities of 673 planetary nebulae (PNe), which was then used to deproject the COBE 2.2 $\mu$m map in order to find $L_d=2.6$ kpc;
\citet{DehnenBinney98}, who optimized dynamical mass models of the MW to a variety of observational constraints available in order to find $L_d/R_0 = 0.3$, and hence $L_d=2.5^\dagger$ kpc; and
\citet{Widrow08}, who used both Bayesian and Markov chain Monte Carlo techniques to match an axisymmetric, equilibrium model of the MW to nine different sets of observational constraints, finding $L_d = 2.80^{+0.23}_{-0.22}$ kpc.
More recently, \citet{Reid14} used Bayesian methods to fit a universal rotation curve to trigonometric parallax and proper motion data for $\sim$100 high-mass star forming regions, finding $L_d=2.34\pm0.18^\dagger$ kpc.  This compares well with our IR estimate, but is inconsistent with our visible and visible+IR estimates at the $\sim$1.4$\sigma$ significance level.

There are a few studies, however, supporting a shorter radial scale length for the Galactic disk.  For instance, \citet{Bienayme99} studied AD by analyzing velocity dispersions measured for $\sim$14,000 stars from the Hipparcos data after segregating them into 20 bins of $B-V$ color, finding $L_d=1.8\pm0.2^\dagger$ kpc.  This estimate is incompatible with the results presented here at the 3--3.5$\sigma$ level.  More recently, \citet{BovyRix13} performed a sophisticated dynamical analysis of $\sim$16,000 G-dwarfs segregated into 43 mono-abundance populations that yielded a mass-weighted average value of $L_d = 2.15\pm0.14$ kpc.  This estimate is in tension with our results at the 1.9--2.6$\sigma$ level if one were to assume that stars trace mass in the disk with a uniform mass-to-light ratio.  However, we note that Hessmann 2016 (submitted) finds that in the DiskMass sample \citep{Bershady10}, mass-to-light ratio varies with radius in disk galaxies, such that the mass scale length is typically 80\% of the disk scale length.  Testing this using our result in combination with the \citet{BovyRix13} mass scale length estimate corresponds to $2.15\pm0.14 / 2.64\pm0.13 = 0.81\pm0.07$, in excellent agreement with that ratio.

On the other hand, there are also studies that found larger values for the Galactic disk scale length.  For instance, \citet{Feast00} applied the AD relation to radial velocity data for semi-regular variables, Mira variables and PNe, yielding a weighted average for the radial density gradient of $R_0/L_d = 2.55\pm0.43$, which corresponds to $L_d = 3.27\pm0.58^\dagger$ kpc.  While this compares fairly well with our visible and visible+IR result, it is inconsistent with our IR result at the 1.3$\sigma$ level.  \citet{McMillan11} took a Bayesian approach to fitting dynamical MW mass models (with 8 free parameters) to a variety of kinematic observational data available, finding $L_d = 3.00\pm0.22$ kpc.  This compares well with our visible result, but is in tension with our IR and visible+IR results at the 1.8$\sigma$ and 1.4$\sigma$ levels, respectively.

\subsection{Comparisons to Multi-band Photometry for Extragalactic Disks} \label{sec:comp_extragalactic}
Due primarily to variations in their star formation histories as a function of radius \citep{deJong96}, spiral galaxies typically become bluer with increasing radius --- the exception being those of bluest integrated color, which are also bluest at their centers \citep[cf.][]{Verheijen97}.  For galaxies like the MW, which ranks among the reddest spirals \citep{Licquia2}, a higher central concentration of older stars causes them to appear more compact when imaged in the IR, whereas younger stars and new star formation extending into the outskirts of the disk causes them to have a broader profile when imaged in the visible.  Hence, scale lengths measured from visible light are usually larger than those measured from IR light, matching the MW results that we find here.

A number of studies have investigated the wavelength dependence of extragalactic scale lengths measured from both optical and IR photometry.  For instance, \citet{Peletier94} performed such an analysis for 37 Sb/c galaxies (closely matching the MW's morphological type; \citealp{dV83}; \citealp{Licquia2}), though with a range of inclinations, and found that extragalactic scale lengths are generally $\sim$1--2$\times$ larger when measured in the visible ($B$-band) compared to the near-IR ($K$-band; see their Figure 4).  \citet{Verheijen97} investigated the correlation of optical-to-IR scale length ratios with disk central surface brightness colors for the disk components of galaxies in the Ursa Major cluster and found similar results (see their Figure 9), with the exception of the bluest galaxies as mentioned above.  While these ratios can reach to much larger values than what we find for the MW here, one must keep in mind that they stem from galaxies with a range of morphological types and, unlike our value, are not corrected for dust attenuation/reddening, which is expected to alter galaxy properties in the optical \citep{Corradi96, Xilouris99, MacArthur04, deGeyter14}.

\citet{deJong96} performed multi-band photometry for 86 face-on disk-dominated galaxies, providing extragalactic data that we may straightforwardly compare to our results.  We have transcribed the data from Table 4 of that paper, which provides scale length measurements in the $B$, $R$, $I$, and $K$ bands, as well as numerical morphological classifications ($T$; \citealt{RC3}).  Restricting to objects that fall into the range $2\leq T\leq6$, which should be similar in morphology to the MW ($T\approx4$), yields a sample of 30 galaxies from which we have measured scale length ratios of $L_d^B/L_d^K=1.17\pm0.34$, $L_d^R/L_d^K=1.14\pm0.26$, and $L_d^I/L_d^K=1.10\pm0.25$ (mean and standard deviation), with median values of 1.05, 1.07, and 1.06, respectively.  Very similar results are found when analyzing the entire sample of galaxies that have $K$-band scale lengths available.  Using our BMM estimates of $L_d$ presented here, we find an optical-to-IR scale length ratio of $2.71/2.51=1.08$ for the MW, making it appear quite typical in this respect compared to other spiral galaxies.

\subsection{A Revised Estimate of the Milky Way's Total Stellar Mass} \label{sec:stellar_mass}
In \citet{Licquia1} we determined updated constraints on the MW's total stellar mass by combining a BMM estimate for the stellar mass in the Galactic bulge+bar with the dynamically-constrained model of the Galactic disk from \citet{BovyRix13}.  As mentioned in the previous section, the scale length estimate yielded from that study is lower than values that we have found here, and is inconsistent with them at the $\sim$2$\sigma$ significance level.  It is likely that systematic differences in what quantity has been measured largely contribute to this tension as the methodology and assumptions used by \citet{BovyRix13} are substantially different from those of the star count-based measurements utilized here.  Furthermore, the \citet{BovyRix13} methodology is not viable for studying external galaxies, whose scale lengths are measured from photometry.  Since a key goal of our work has been comparing the properties of the MW to those of extragalactic objects, it is beneficial to produce a revised estimate of the stellar mass of the Galaxy by developing a model of the disk that utilizes the light-based scale length estimates determined in this study.

\subsubsection{Updating the Disk Model Assumptions} \label{sec:disk_mass_assumptions}
We begin by making several adjustments to the assumptions that went into the the original disk model described in \S4.1 of \citet{Licquia1}.  From Equation 17 of that paper, the stellar mass of the MW's exponential disk, given its radial scale length, $L_d$, the radius of the Sun from the Galactic center, $R_0$, and the \emph{mean} surface density of stellar material at $R_0$, $\bar{\Sigma}_\star(R_0)$, is determined by 
\begin{equation} 
\textrm{M}_\star^\text{D} = 2\pi\bar{\Sigma}_\star(R_0) L_d^2 \exp(R_0 / L_d). \label{eq:diskmass} 
\end{equation}
Note that here we have made a slight notation change by using $\bar{\Sigma}_\star(R_0)$ in place of $\Sigma_\star(R_0)$; this is to remind the reader that Equation \eqref{eq:diskmass} assumes the stellar disk to be axisymmetric about the Galactic center.  In our past work, just as the vast majority of previous studies have done, we assumed that the \emph{locally} measured surface density --- i.e., the column density at the location of the Sun --- is representative of the mean value.  However, the MW's spiral structure certainly produces variations in the observed $\Sigma_\star(R_0,\phi)$ as one varies $\phi$.  For reasons that will be explained in the following section, we no longer make this assumption and hence $\bar{\Sigma}_\star(R_0)$ represents the stellar surface density at $R=R_0$ averaged over all $\phi$.  For further clarity, hereafter we denote the locally measured surface density --- i.e., the observed value for the Solar neighborhood --- as $\Sigma_\star(R_0, \phi_0)$.

Next, since near-IR light is expected to most closely trace the distribution of stellar mass, we now adopt the BMM result from analyzing IR estimates only in this paper for the disk scale length, which corresponds to $L_d = 0.12(R_0 - 8.33\,\text{kpc})+2.51^{+0.15}_{-0.13}$ kpc when including the covariance with $R_0$ that we have found in \S\ref{sec:r0dependence}.  Since this is entirely independent of the surface mass density, this nullifies the covariance between $L_d$ and $\Sigma_\star(R_0, \phi_0)$ that we needed to handle in the \citet{BovyRix13} model (see Table 3 of \citealt{Licquia1}).  Consequentially, we adopt the dynamically-derived estimate of $\Sigma_\star(R_0, \phi_0) = 34.8\pm4.3$ \massunits\ pc$^{-2}$, which we arrived at before accounting for any covariance with $L_d$ in that paper.  As discussed therein, this value matches very well with estimates from photometric techniques \citep[cf.][]{Flynn06, Hessman15, McKee15}.

Furthermore, we no longer enforce a linear scaling relationship between $\textrm{M}_\star^\text{D}$ and values of $R_0$ differing from 8 kpc, as reported by \citet{BovyRix13}; instead, we simply allow for the exponential dependency that comes naturally with Equation \eqref{eq:diskmass}.  We continue to use our original prior on $R_0$, corresponding to $8.33\pm0.35$ kpc, as it is consistent with essentially all recent measurements in the literature, and hence reasonably describes our current knowledge of this parameter.  %We note that while we have used this value to standardize a subset of the IR scale length measurements analyzed here (as indicated in Table \ref{table:HB_data}), our results display only a weak correlation for these parameters; testing for variations in our choice of $R_0$ revealed that $L_d\propto\sim0.1R_0$.  We find that the results from our revised disk model are entirely unchanged whether we include this covariance or not, and so for simplicity we treat $R_0$ and $L_d$ as independent hereafter.

\subsubsection{Correcting for Local Density Variations} \label{sec:density_var}
As mentioned above, it is important to note that Equation \eqref{eq:diskmass} is obtained by integrating the surface density profile for an exponential disk under the assumption that it is axisymmetric and smooth.  Most prior studies of the Galactic disk mass, including our own past work, simply took $\Sigma_\star(R_0, \phi_0)$ to be representative of the mean value and plugged this in for $\bar{\Sigma}_\star(R_0)$ in Equation \eqref{eq:diskmass}.  Given that the MW is known to be a spiral galaxy, a more realistic justification of this is to assume that density perturbations due to spiral structure are negligible compared to the smooth, underlying exponential distribution of stars, or that $\Sigma_\star(R_0, \phi_0)$ is unperturbed from $\bar{\Sigma}_\star(R_0)$.  There is substantial evidence, however, suggesting that both of these assumptions are inaccurate.  

Broadly speaking, the Sun is believed to lie in a region of rarefaction within the MW's spiral structure, roughly midway between the Sagittarius and Perseus arms (see, e.g., D\&S01, \citealp{Quillen02}; \citealp{Xu13}).  Assuming that this is true in detail would indicate that the local surface density of stellar material must be rescaled to use it in a smooth model for the disk.  However, it is also worth noting that the Sun lies nearby to or within a branch of star-forming regions collectively called the ``Orion Spur'' or ``Local Arm'' \citep[][and references therein]{Vallee95, Xu13}, a secondary feature in the spiral structure that is less prominent than the main arms, but also resides within a $\sim$100--200 pc wide cavity of relatively sparse cold and neutral interstellar gas known as the ``Local Bubble'' \citep{CoxReynolds87, Lallement14}.  This makes it murky at best as to what the true correction factor should be, and this is likely why previous MW models have typically neglected this effect altogether (\citealt{Flynn06} is one example where the authors do account for this effect by incorporating a 10\% correction for spiral enhancement).  Therefore, we next explore both the Galactic and extragalactic data available in order to get a handle on the size of this correction.

The effect of density perturbations is generally parameterized by the spiral arm amplitude, $A$, which denotes the fractional increase in the surface density (or brightness) compared to the mean.  In many cases, authors report the arm-to-interarm ratio, which is related to the spiral amplitude by $\Sigma_\text{arm}/\Sigma_\text{interarm} = (1+A)/(1-A)$.  For instance, D\&S01 produced a model of the dust, stars, and spiral arms in the MW's disk from \textit{COBE}/DIRBE data, which required placing the Sun within a gap in the spiral features, consistent with the picture described above.  They found arm-to-interarm ratios of 1.2 and 1.32 in the $J$- and $K$-band, respectively, corresponding to values of $A=0.09$ and 0.14.  \citet{Liu12} investigated the extinction and radial velocity dispersion and distribution of red clump stars in the Perseus arm, finding $\Sigma_\text{arm}/\Sigma_\text{interarm} =$ 1.3--1.5, corresponding to $A=$ 0.13--0.20.  Most recently, P13 fit the O\&L93 star count model of the Galaxy, which accounts for spiral structure (see Table 1), to all-sky 2MASS data, yielding an estimate of $\Sigma_\text{arm}/\Sigma_\text{interarm} = 2.0^{+0.6}_{-0.8}$, corresponding to $A=0.33^{+0.11}_{-0.24}$.

\citet{Grosbol04} obtained deep $K$-band surface photometry for 54 normal spiral galaxies and analyzed their spiral structure using Fourier techniques.  This resulted in a skewed distribution of spiral amplitudes ranging from $\sim$0--0.5 with a median value of $\sim$0.2, which we have measured from data in their Table 2.  \citet{Elmegreen11} investigated the spiral arm properties of 46 Sa--Sm galaxies with varying characteristics from their 3.6 $\mu$m images taken in the \textit{Spitzer} Survey of Stellar Structure in Galaxies (S$^4$G).  The authors measured arm and interarm surface brightnesses and tabulated their ratios converted to a magnitude in their Table 2.  We have converted these values back to arm-to-interarm ratios based on Equation 3 of that study and find that they span the range of 1.3--4.0 (with one large outlier at 8.3 that we disregard), correspond to $A=$ 0.13--0.6.  Lastly, \citet{Kendall15} analyzed the spiral structure of 13 galaxies from the \textit{Spitzer} Infrared Nearby Galaxies Survey (SINGS).  Column 3 of their Table 2 provides spiral amplitudes measured at 3.6 $\mu$m, which span the range of $\sim$0.1--0.4.

In order to derive an estimate for the spiral amplitude in the MW we statistically combine these Galactic and extragalactic measurements, treating each as a Gaussian distribution with its $\pm$1$\sigma$ region matching the range of values detailed above.  Calculating the unweighted mean yields $A=0.225\pm0.095$, whereas the inverse variance-weighted mean yields $A=0.158\pm0.024$, and the median value is 0.193.  Considering these metrics, we conservatively adopt $A=0.2\pm0.1$ as the spiral amplitude of the Galaxy.  Assuming that the Sun lies in a density trough, the appropriate correction factor to include in our axisymmetric disk model is $\Sigma_\text{mean}/\Sigma_\text{interarm} = (1-A)^{-1} = 1.25^{+0.18}_{-0.14}$ (median and 1$\sigma$ range).  However, as we mentioned above, there is considerable uncertainty about the net local density perturbation, and it is certainly possible that the local surface density is representative of the Solar circle.  We therefore adopt a correction factor, $C \equiv \bar{\Sigma}_\star(R_0) / \Sigma_\star(R_0, \phi_0)$, for the density variation at the Sun's location with an additional amount of uncertainty, such that a non-correction is included well within the 2$\sigma$ significance level, corresponding to $C = 1.25^{+0.29}_{-0.20}$ (this is equivalent to $(1-A)^{-1}$ after augmenting the errors on $A$ by 50\% compared to the value derived above.)  We then multiply $C$ by $\Sigma_\star(R_0, \phi_0)$ to obtain $\bar{\Sigma}_\star(R_0) = 43.4^{+11.5}_{-8.4}$ \massunits\ pc$^{-2}$.

%We therefore have added an additional amount of uncertainty to this correction value, such that a non-correction is included well within the 2$\sigma$ significance level, corresponding to $(1-A)^{-1} = 1.25^{+0.29}_{-0.20}$, or equivalently $A=0.20\pm0.15$.

\subsubsection{Revised Stellar Mass Results} \label{sec:disk_mass_assumptions}
To summarize the above, we have developed a new model of the Galactic disk where the stellar density follows a smooth, axisymmetric exponential profile with an underlying spiral arm structure characterized by a density amplitude of $A=0.2\pm0.1$.  We assume that the Sun lies at a Galactocentric radius of $R_0=8.33\pm0.35$ kpc, but also in a rarefaction region of the spiral structure, roughly midway between the Sagittarius and Perseus arms, and so local measures of the stellar surface density are not representative of the mean stellar surface density along the annulus at $R=R_0$ (i.e., averaged over all $\phi$).  The appropriate correction factor is $\Sigma_\text{mean}/\Sigma_\text{interarm} = (1-A)^{-1} = 1.25^{+0.18}_{-0.14}$; however, given the uncertainties about the net local density perturbation described above, we conservatively adopt a correction factor for the Sun's position with an augmented error bar given by $C \equiv \bar{\Sigma}_\star(R_0) / \Sigma_\star(R_0, \phi_0) = 1.25^{+0.29}_{-0.20}$.  From our previous work we adopt a dynamically-derived estimate of the local stellar surface density of $\Sigma_\star(R_0, \phi_0) = 34.8\pm4.3$ \massunits\ pc$^{-2}$, and by multiplying this with $C$ we obtain an estimate for the mean value at $R_0$ given by $\bar{\Sigma}_\star(R_0) = 43.4^{+11.5}_{-8.4}$ \massunits\ pc$^{-2}$.  Given that the exponential distribution of stars should be well traced by near-IR light, we adopt the IR photometric scale length estimate derived here of $L_d = 0.12(R_0 - 8.33\,\text{kpc})+2.51^{+0.15}_{-0.13}$ kpc, which includes the covariance with $R_0$.  Lastly, the stellar mass of the Galactic disk corrected for spiral arm density variations is calculated by plugging these values into Equation \eqref{eq:diskmass}. 
 
Using this model, we have next repeated the analysis from \citet{Licquia1} in order to produce a revised estimate of the total stellar mass, $\textrm{M}_\star$.  As described in \S4.2 of that paper, this entails using Monte Carlo techniques to produce model-consistent realizations of the disk mass and the BMM estimate of the bulge+bar mass (the stellar halo contribution is assumed smaller than the uncertainties in $\textrm{M}_\star^\text{D}$ and hence negligible; e.g., \citealp{Bell08}), the results of which are summed to obtain the posterior probability distribution for $\textrm{M}_\star$.  We find that while the bulge+bar mass estimate remains $0.91\pm0.07\times10^{10}$ \massunits, the posteriors for the disk and total stellar mass are described by $\textrm{M}_\star^\text{D} = 4.8^{+1.5}_{-1.1}\times10^{10}$ \massunits\ and $\textrm{M}_\star = 5.7^{+1.5}_{-1.1}\times10^{10}$ \massunits\ (median and 1$\sigma$ range), or equivalently $\log(\textrm{M}_\star^\text{D}/\textrm{M}_\odot) = 10.68\pm0.12$ and $\log(\textrm{M}_\star/\textrm{M}_\odot) = 10.75\pm0.10$, which are consistent with our previous results at the 0.2$\sigma$ level.  Our revised model also corresponds to a bulge-to-total mass ratio of $0.16\pm0.03$ and a specific star formation rate of $\mathrm{\dot{M}}_\star/\mathrm{M}_\star=2.89^{+0.80}_{-0.67}\times10^{-11}$ yr$^{-1}$, or equivalently $\log(\mathrm{\dot{M}}_\star/\mathrm{M}_\star/\textrm{yr}^{-1})=-10.54\pm0.11$, when combined with the star formation rate estimate of $\mathrm{\dot{M}}_\star=1.65\pm0.19$ \sfrunits\ from our original paper.

We refer the reader to Appendix \ref{sec:updated_props} for a convenient tabulation of the properties derived in this section.  There one will also find updated estimates for the photometric properties of the MW, which we have produced by entirely reperforming the analyses of \citet{Licquia2} using the model of the Galactic disk and hence total stellar mass derived here.  We note that differences between these values and the original ones published in that paper are marginal compared to the uncertainties, as expected given the small change in the stellar mass estimate here compared to our original estimate in \citet{Licquia1}.  Nevertheless, the values tabulated in Appendix \ref{sec:updated_props} are all self-consistent and uniformly reflect our best-to-date model of the MW.

\section{Summary and Conclusions} \label{sec:summary}
In this study, we have set out to determine a combined, robust estimate of the scale length of the Galactic disk, $L_d$, measured at visible and IR wavelengths, given the large array of data available in the literature.  Upon thoroughly investigating the previous estimates of $L_d$ (see Table 1), we find that the set of Galactic models that are employed display as much variety as the observational datasets they are optimized to match, typically containing around a dozen free parameters.  Aside from the wide assortment of model assumptions involved, given that measurements of $L_d$ are produced from fitting the smooth underlying structure of the disk, these estimates are also susceptible to systematic error due to undetected substructures present in the true distribution of stars along any particular line of sight through the Galaxy (J08).  As a result of these variations in methodology and data, the estimates we have collected fall anywhere in the range $2\lesssim L_d\lesssim6$ kpc.  This is comparable to the dynamical range of scale lengths measured for other galaxies of similar mass to the MW in the local Universe ($1\lesssim L_d\lesssim10$ kpc; cf. Licquia et al. 2016, in preparation), and hence an improved determination of the scale length will also improve our understanding of how the MW fits amongst the broader population of galaxies. 

Foregoing less sophisticated meta-analysis techniques (e.g., the inverse variance-weighted mean; IVWM), we have opted for a robust analysis method that has proven powerful in many applications inside and outside of astronomy.  More specifically, we have produced our results by statistically combining the literature estimates using a Bayesian mixture-model (BMM) approach, which allows us to account for the possibility that one or more of these estimates have not properly accounted for all statistical or systematic errors.  Through Monte Carlo techniques we have ensured that all the estimates used are rescaled to reflect current knowledge of the Sun's distance from the Galactic center, $R_0$.  Lastly, we have implemented a Bayesian model averaging (BMA) technique to obtain the posterior for $L_d$ marginalized over all the bad-measurement models we have investigated, taking into account both goodness-of-fit and model complexity.  Ultimately, we find the Galactic scale length to be $L_d = 2.71^{+0.22}_{-0.20}$ kpc for visible starlight, $L_d = 2.51^{+0.15}_{-0.13}$ kpc for IR starlight, and $L_d = 2.64\pm0.13$ kpc when integrating visible and IR starlight measurements (see \S\ref{sec:r0dependence} for discussion on how these results depend on $R_0$).

In Table \ref{table:HB_results} we have listed a full summary of results from our bad-measurement models, the result from applying BMA, as well as the IVWM for comparison.  We have demonstrated in \S\ref{sec:robust} that our results are robust to varying many of the assumptions we have made in our analyses, and in \S\ref{sec:comp_extragalactic} that they are consistent with passband-to-passband variations measured for external disks.  We have also used our results to revise the estimate of the MW's total stellar mass from \citet{Licquia1} in \S\ref{sec:stellar_mass}.  Using the IR scale length measurement found here, we find that the mass of the stellar disk is $4.8^{+1.5}_{-1.1}\times10^{10}$ \massunits.  Combining this with the BMM estimate for the bulge+bar mass in a model-consistent manner using the framework of \citet{Licquia1}, we have determined the MW's total stellar mass to be $5.7^{+1.5}_{-1.1}\times10^{10}$ \massunits.  For convenience, we have compiled in Appendix \ref{sec:updated_props} several tables displaying the updated constraints we have produced for the structural and mass properties of the MW, as well as updates to the results from \citet{Licquia2} using the stellar mass derived here.

The remaining estimates of $L_d$ in the literature we have compared our results to are generically constrained by modeling stellar kinematic observations, and hence describe the radial distribution of \emph{total} mass in the Galaxy.  Nevertheless, the majority of such estimates compare well with the values we have presented in this study based on visible/IR starlight, though there are a few that are in significant tension, favoring instead values of $L_d$ far below or above our constraints.  It is beyond the scope of this paper to comment on whether a dynamically-constrained $L_d$ is a fair comparison with those based on starlight, but this agreement adds some credence to our stellar scale length estimates.  Interestingly, it appears that dynamical estimates are as prevalent \emph{and} as disparate as photometric estimates, and astronomers in need of adopting a value from the literature would likely benefit from performing the type of BMM analysis we have employed here to that set as well.

In totality, the results of this study, in combination with those of our previous works which we have updated here, provide a much improved comprehensive picture of the MW.  More specifically, we have determined tight constraints on a variety of the Galaxy's global properties, including its total stellar mass, star formation rate, photometric disk scale length, and optical luminosity and color index, using methods that either circumvent or correct for many of the major systematics that have traditionally affected them.  Moreover, all of these values have been produced from a single, consistent model of the MW that reflects our best-to-date knowledge its structural parameters, and which rests upon the same basic assumptions that are used for studying extragalactic objects.  All of this work culminates in a newfound ability to assess accurately how the properties of our Galaxy compare to scaling relations found for external spiral galaxies.  In a companion paper to this one (Licquia et al. 2016; in preparation), we will present new comparisons of the MW to both the Tully-Fisher relation as well as 3-dimensional luminosity-velocity-size relations for other massive spiral galaxies in order to assess how our Galaxy truly fits in a variety of extragalactic contexts.
\clearpage

\section*{Acknowledgements}
It is a pleasure to thank Brett Andrews for reading an early draft of this paper and providing valuable feedback.  We also thank Chad Schafer, Matt Bershady, and Rick Hessman for a number of helpful discussions, as well as the anonymous referee for leading us to several improvements in this work.  TCL and JAN are supported by the National Science Foundation (NSF) through grant NSF AST 08-06732.  TCL is also supported as a PITT PACC fellow.

\bibliographystyle{apj}
\bibliography{mw_ld}
\normalsize
\appendix
\section{Updated Estimates of Milky Way Properties} \label{sec:updated_props}
The following tables provide a comprehensive summary of the Milky Way properties derived herein, as well as those derived in \citet{Licquia1} and \citet{Licquia2} after updating them to the model of the stellar disk from \S\ref{sec:stellar_mass} of this study, which utilizes our estimate for the IR photometric disk scale length and accounts for local density variations due to spiral structure.
\FloatBarrier

\begin{deluxetable*}{lccc}
\tablewidth{0pt}
\tablenum{5}\label{table:struct_props}
\tablecaption{Structural Properties of the Milky Way}
\tablehead{Parameter & Optimal Value $\pm$1$\sigma$ & Units & Source}
\startdata
$A$ & $0.2\pm0.1$ & & derived in this study \\
$R_0$ & $8.33\pm0.35$ & kpc & \citet{Gillessen} \\
$L_d$ & $2.51^{+0.15}_{-0.13}+0.12(R_0/\text{kpc} - 8.33)$ & kpc & derived in this study \\
$\Sigma_\star(R_0, \phi_0)$ & $34.8\pm4.3$ & \massunits\ pc$^{-2}$ & \citet{Licquia1} \\
$C$ & $1.25^{+0.29}_{-0.20}$ & & derived in this study \\
$\bar{\Sigma}_\star(R_0)$ & $43.4^{+11.5}_{-8.4}$ & \massunits\ pc$^{-2}$ & derived in this study
\enddata
\tablecomments{A description of each parameter is as follows: $A$ is the spiral amplitude of the Milky Way disk, $R_0$ is the radius of the Sun from the Galactic center, $L_d$ is the photometric disk scale length measured from IR starlight, $\Sigma_\star(R_0, \phi_0)$ is the surface density of stellar material (main sequence stars plus stellar remnants, but not brown dwarfs) at the Sun's location, and $C$ is the ratio between the \emph{mean} surface density at $R=R_0$, denoted by $\bar{\Sigma}_\star(R_0)$, and $\Sigma_\star(R_0, \phi_0)$.  See \S\ref{sec:stellar_mass} for discussion of these estimates.}
\end{deluxetable*}

\begin{deluxetable*}{lccc}
\tablewidth{0pt}
\tablenum{6}\label{table:mass_props}
\tablecaption{Mass Properties of the Milky Way}
\tablehead{Parameter & Optimal Value $\pm$1$\sigma$ & Units & Source}
\startdata
$\textrm{M}_\star^\text{B}$ & $0.91\pm0.07\times10^{10}$ & \massunits & derived in this study \\
$\textrm{M}_\star^\text{D}$ & $4.8^{+1.5}_{-1.1}\times10^{10}$ & \massunits & derived in this study \\
$\textrm{M}_\star$ & $5.7^{+1.5}_{-1.1}\times10^{10}$ & \massunits & derived in this study \\
$B/T$ & $0.16\pm0.03$ & & derived in this study \\
$\mathrm{\dot{M}}_\star$ & $1.65\pm0.19$ & \sfrunits & \citet{Licquia1} \\
$\mathrm{\dot{M}}_\star / \textrm{M}_\star$ & $2.89^{+0.80}_{-0.67}\times10^{-11}$ & yr$^{-1}$ & derived in this study
\enddata
\tablecomments{A description of each parameter is as follows: $\textrm{M}_\star^\text{B}$ is the stellar mass of the central bulge+bar, $\textrm{M}_\star^\text{D}$ is the stellar mass of the disk, $\textrm{M}_\star$ is the total stellar mass, $B/T$ is the mass bulge-to-total ratio, $\mathrm{\dot{M}}_\star$ is the global star formation rate, and $\mathrm{\dot{M}}_\star / \textrm{M}_\star$ is the specific star formation rate.  Note that the mass of the stellar halo is assumed negligible (it is much smaller than the uncertainties in $\textrm{M}_\star^\text{D}$; e.g., \citealp{Bell08}) in this model, such that $\textrm{M}_\star \equiv \textrm{M}_\star^\text{B} + \textrm{M}_\star^\text{D}$. See \S\ref{sec:disk_mass_assumptions} for discussion of these estimates.}
\end{deluxetable*}

\begin{deluxetable*}{lcclc}
\tablenum{7}\label{table:updated_props1}
\tablewidth{0pt}
\tablecaption{Photometric Properties for the Milky Way: Rest--frame $z$=0 Passbands}
\tablehead{Absolute & Optimal Value $\pm$1$\sigma$ & & Color & Optimal Value $\pm$1$\sigma$ \\
Magnitude & (mag) & & Index & (mag)}
\startdata
$^0\!M_u - 5\log h$ & $-19.15_{-0.47}^{+0.55}$ & $\quad\quad$ &$^0(u-r)$ & $2.029_{-0.150}^{+0.153}$ \\
$^0\!M_g - 5\log h$ & $-20.33_{-0.43}^{+0.42}$ & & $^0(u-g)$ & $1.349_{-0.092}^{+0.107}$ \\
$^0\!M_r - 5\log h$ & $-20.97_{-0.40}^{+0.37}$ & & $^0(g-r)$ & $0.678_{-0.057}^{+0.069}$ \\
$^0\!M_i - 5\log h$ & $-21.24_{-0.38}^{+0.37}$ & & $^0(r-i)$ & $0.294_{-0.046}^{+0.052}$ \\
$^0\!M_z - 5\log h$ & $-21.53_{-0.39}^{+0.36}$ & & $^0(i-z)$ & $0.288_{-0.041}^{+0.042}$ \\
\\[-1ex]
$^0\!M_U - 5\log h$ & $-20.00_{-0.47}^{+0.59}$ & & $^0(U-V)$ & $0.879_{-0.125}^{+0.150}$ \\
$^0\!M_B - 5\log h$ & $-20.05_{-0.45}^{+0.41}$ & & $^0(U-B)$ & $0.143_{-0.071}^{+0.082}$ \\
$^0\!M_V - 5\log h$ & $-20.71_{-0.40}^{+0.39}$ & & $^0(B-V)$ & $0.740_{-0.056}^{+0.065}$ \\
$^0\!M_R - 5\log h$ & $-21.23_{-0.39}^{+0.39}$ & & $^0(V-R)$ & $0.540_{-0.041}^{+0.044}$ \\
$^0\!M_I - 5\log h$  & $-21.81_{-0.38}^{+0.38}$ & & $^0(R-I)$  & $0.594_{-0.049}^{+0.050}$
 \enddata
 \tablecomments{These values are found by reperforming the analysis of \citet{Licquia2} using the updated model the Galactic disk and hence total stellar mass derived in \S\ref{sec:stellar_mass}.  The corresponding changes in these values are marginal compared to the uncertainties.}
 \end{deluxetable*} 

\begin{deluxetable*}{lcclc}
\tablenum{8}\label{table:updated_props2}
\tablewidth{0pt}
\tablecaption{Photometric Properties for the Milky Way: Rest--frame $z$=0.1 Passbands}
\tablehead{ Absolute & Optimal Value $\pm$1$\sigma$ & & Color & Optimal Value $\pm$1$\sigma$ \\
Magnitude & (mag) & & Index & (mag)}
\startdata
$^{0.1}\!M_u - 5\log h$ & $-18.84_{-0.50}^{+0.56}$ &$\quad\quad$ &$^{0.1}(u-r)$ & $2.187_{-0.164}^{+0.193}$ \\
$^{0.1}\!M_g - 5\log h$ & $-20.05_{-0.46}^{+0.46}$ & &$^{0.1}(u-g)$ & $1.411_{-0.113}^{+0.121}$ \\
$^{0.1}\!M_r - 5\log h$  & $-20.75_{-0.40}^{+0.38}$ & & $^{0.1}(g-r)$ & $0.777_{-0.065}^{+0.078}$ \\
$^{0.1}\!M_i - 5\log h$  & $-21.13_{-0.40}^{+0.36}$ & & $^{0.1}(r-i)$  & $0.389_{-0.042}^{+0.047}$ \\
$^{0.1}\!M_z - 5\log h$ & $-21.38_{-0.39}^{+0.35}$ & & $^{0.1}(i-z)$ & $0.272_{-0.048}^{+0.047}$ \\
\\[-1ex]
$^{0.1}\!M_U - 5\log h$ & $-20.08_{-0.50}^{+0.54}$ & & $^{0.1}(U-V)$ & \phs$0.592_{-0.136}^{+0.161}$ \\
$^{0.1}\!M_B - 5\log h$ & $-19.93_{-0.46}^{+0.46}$ & & $^{0.1}(U-B)$ & $-0.021_{-0.092}^{+0.097}$ \\
$^{0.1}\!M_V - 5\log h$ & $-20.44_{-0.41}^{+0.42}$ & & $^{0.1}(B-V)$ & \phs$0.620_{-0.063}^{+0.073}$ \\
$^{0.1}\!M_R - 5\log h$ & $-20.95_{-0.39}^{+0.38}$ & & $^{0.1}(V-R)$ & \phs$0.516_{-0.042}^{+0.047}$ \\
$^{0.1}\!M_I - 5\log h$  & $-21.57_{-0.38}^{+0.38}$ & & $^{0.1}(R-I)$  & \phs$0.634_{-0.045}^{+0.050}$
 \enddata
\tablecomments{These values are found by reperforming the analysis of \citet{Licquia2} using the updated model the Galactic disk and hence total stellar mass derived in \S\ref{sec:stellar_mass}.  The corresponding changes in these values are marginal compared to the uncertainties.}
\end{deluxetable*} 

\begin{deluxetable*}{lccccc}
\tablenum{9}\label{table:M2L}
\tablewidth{0pt}
\tablecaption{Global Stellar Mass--to--light Ratios for the Milky Way}
\tablehead{Rest-- & \multirow{2}{*}{$\Upsilon^\star_u$} & \multirow{2}{*}{$\Upsilon^\star_g$} & \multirow{2}{*}{$\Upsilon^\star_r$} & \multirow{2}{*}{$\Upsilon^\star_i$} & \multirow{2}{*}{$\Upsilon^\star_z$} \\
frame &&&&&}
\startdata
$z$=0 & $1.84^{+1.20}_{-0.76}$ & $1.93^{+0.72}_{-0.64}$ & $1.64^{+0.52}_{-0.51}$ & $1.41^{+0.41}_{-0.42}$ & $1.10^{+0.32}_{-0.32}$ \\
\\
$z$=0.1 & $1.71^{+1.29}_{-0.79}$ & $1.87^{+0.86}_{-0.63}$ & $1.82^{+0.67}_{-0.57}$ & $1.51^{+0.53}_{-0.44}$ & $1.26^{+0.38}_{-0.37}$ \\[1ex]
\hline
\hline
\\[-1.4ex]
Rest-- & \multirow{2}{*}{$\Upsilon^\star_U$} & \multirow{2}{*}{$\Upsilon^\star_B$} & \multirow{2}{*}{$\Upsilon^\star_V$} & \multirow{2}{*}{$\Upsilon^\star_R$} & \multirow{2}{*}{$\Upsilon^\star_I$} \\
frame &&&&& \\
\hline
\\[-1.4ex]
$z$=0 & $1.81^{+1.12}_{-0.75}$ & $1.85^{+0.80}_{-0.63}$ & $1.84^{+0.66}_{-0.59}$ & $1.59^{+0.56}_{-0.47}$ & $1.28^{+0.44}_{-0.37}$ \\
\\
$z$=0.1 & $1.76^{+1.26}_{-0.81}$ & $1.81^{+0.84}_{-0.65}$ & $1.92^{+0.84}_{-0.60}$ & $1.72^{+0.60}_{-0.53}$ & $1.41^{+0.45}_{-0.41}$ 
\enddata
\tablecomments{All values are expressed in units of $\massunits/L_\odot$ and are found by reperforming the analysis of \citet{Licquia2} using the updated model the Galactic disk and hence total stellar mass derived in \S\ref{sec:stellar_mass}.  The corresponding changes in these values are marginal compared to the uncertainties.}
\end{deluxetable*} 

\end{document}